\documentclass[journal]{IEEEtran}
\usepackage{tabularx}

\usepackage{comment}
\usepackage{amsmath}
\usepackage{amssymb}
\usepackage{dblfloatfix} 
\usepackage{physics}
\usepackage{color}

\usepackage{setspace}
\usepackage{booktabs}
\usepackage{pifont}

\usepackage{amsmath}
\DeclareMathOperator*{\argmax}{arg\,max}

\usepackage{float}
\usepackage{subfloat}

\ifCLASSINFOpdf
   \usepackage[pdftex]{graphicx,psfrag}
   \graphicspath{{../pdf/}{../jpeg/}}
   \DeclareGraphicsExtensions{.pdf,.jpeg,.png}
\else
   \usepackage[dvips]{graphicx}
   \graphicspath{{../eps/}}
   \DeclareGraphicsExtensions{.eps}
\fi

\usepackage{subfigure}

\usepackage{circuitikz}

\usepackage{tikz}

\begin{document}
%
\title{A Chirp Spread Spectrum Modulation Scheme for Robust Power Line Communication}

\author{Stephen~Robson       
        and~Manu~Haddad,~\IEEEmembership{Member,~IEEE}
\thanks{S. Robson and A. Haddad are with the Advanced High Voltage Engineering Research Centre (AHIVE)
, Cardiff University, UK. e-mail: robsons1@cardiff.ac.uk}
\thanks{Manuscript received June 19, 2021; revised:.}}

%
%

\markboth{}%
{Shell \MakeLowercase{\textit{et al.}}: Bare Demo of IEEEtran.cls for Journals}
%



\maketitle

\begin{abstract}
This paper proposes the use of a LoRa like chirp spread spectrum physical layer as the basis for a new Power Line Communication modulation scheme suited for low-bandwidth communication. It is shown that robust communication can be established even in channels exhibiting both extreme multipath interference and low SNR (-40dB), with synchronisation requirements significantly reduced compared to conventional LoRa. ATP-EMTP simulations using frequency dependent line and transformer models, and simulations using artificial Rayleigh channels demonstrate the effectiveness of the new scheme in providing load data from LV feeders back to the MV primary substation. We further present experimental results based on a Field Programmable Gate Array hardware implementation of the proposed scheme.
\end{abstract}


%
\IEEEpeerreviewmaketitle

\section{Introduction}
%
%
%
%

\IEEEPARstart{D}{istribution} Network Operators (DNOs) already deploy a wide range of communication technologies to support the move towards smart grids. Advances in the standardisation of narrowband Power Line Communication (PLC) solutions such as Prime and G3-PLC, and growing options in long range wireless communication (i.e. LoRa) have only added to the options in the last decade. But there remains an unmet requirement for low-cost and robust communication of, for example, load data from secondary substations. The problem is amplified by the sheer number of required monitoring points (typically tens of thousands in a large regional distribution network) and the fact that secondary substations are often located in rural areas with limited access to conventional wired or wireless communication infrastructure.

Previous attempts at providing widescale communication across large distribution networks have tended to focus on the use of conventional wired media (i.e. ethernet), wireless solutions (LoRa, GSM) or PLC. 

Narrowband PLC solutions such as Prime \cite{primestandard} and G3-PLC \cite{g3plcstandard} are now firmly established on Low-Voltage (LV) networks, and are often deployed in automatic meter reading (AMR) applications and increasingly in support of other smart grid services. Over longer distances and across voltage levels, these technologies struggle to cope with the increased attenuation and extreme multipath conditions associated with transmission through transformers and Medium Voltage (MV) networks \cite{robson2016}. New technologies are required in this space. 

This paper proposes a PLC modulation scheme based on the Chirp Spread Spectrum (CSS) scheme of the recently standardised LoRa physical layer \cite{lorasemtech}. The modification is designed to combat the two major problems of extreme multipath and low SNR. The former is resolved by subdividing the LoRa symbol into a reduced set, thereby containing the multipath energy into a single symbol. The latter is resolved through the use of statistical averaging of the modified signal over consecutive symbols. This trading off of data rate for performance makes possible a communication scheme in which many LV feeder monitoring devices can communicate back to a primary substation at timescales of several seconds or minutes.

\section{Background}
\subsection{Requirements for Robust Communication on the LV-MV channel}
The communication channel linking the LV and MV parts of a distribution network is characterised by extreme multipath conditions ($\sigma_{rms}$ = 10's to 100's of $\mu$s). The main contributing factor to this is not the attenuation of the power line itself, rather it is a result of delayed versions of the signal reaching the receiver from many different paths. On the MV network, the typical lengths of the line become much larger than the wavelength of the narrowband PLC signal, and lines are terminated by open circuits or transformers with large reflection coefficients. Therefore, much of the signal energy remains in the power line until it dissipates. Empirical measurements of the RMS delay spread on MV distribution networks is in the tens of $\mu$s \cite{10kv}. In contrast, the RMS delay spread on LV networks is less than 10 $\mu$s \cite{lv_delay}. Therefore, a robust communication scheme suited to this environment must accommodate extremely high RMS delay spreads, far beyond what is typical in LV and conventional wireless systems.

When considering cross-network (LV-MV) transmission, for example in a system which relays load information from a secondary to a primary substation, a second major problem emerges. Though it has been demonstrated that PLC signals in the narrowband range (15-500 kHz) can indeed propagate through transformers, the attenuation is large. Empirical measurements show an average 35 dB attenuation with a high degree of frequency selectivity \cite{transformer_measurement}. The SNR penalty imposed by this scale of attenuation renders existing narrowband PLC technologies unusable. The situation is further exacerbated by regulatory limits on transmit power on power lines. Therefore, reliable inter-transformer communication is only possible with communication schemes that can work at low SNRs.

The dual problem of extreme multipath and high attenuation makes the design of a communication system for this channel extremely challenging. Recently, the emerging LoRa standard was proposed for PLC communication \cite{lorarobson}, and then for time disemmination \cite{loratoa}\cite{robson_toa}. LoRa has several favourable properties, including excellent receiver sensitivity and low power. However, in its raw form, it performs poorly in severe multipath. Here, we exploit the unique properties of LoRa, with a few key modifications, for robust performance in the low SNR and high multipath regime.

\subsection{The LoRa physical layer}

The mathematical basis underpinning the LoRa physical layer has been studied extensively in several recent works \cite{loramath}\cite{lora_2}\cite{approx}. LoRa transmits symbols as frequency shifted chirps. With a bandwidth of $B = \frac{1}{T}$, the transmitted symbol, $w_k$ is defined as:

\begin{equation}
 w_k(nT) = \sqrt{\frac{E_s}{2^{SF}}} e^{j2\pi \cdot (k+n) \mod 2^{SF} \cdot \frac{n}{2^{SF}}}   \label{eqn:2}
    \end{equation}


The above equation describes a series of $n=0,1,2\dots 2^{SF-1}$ consecutive samples forming a LoRa symbol. $SF \in \{7,8\dots,12\}$ is the so called Spreading Factor, which determines the number of transmitted samples per LoRa symbol. $k\in 0,1,2\dots 2^{SF-1}$ is the transmitted symbol and $E_s$ is the symbol energy. It has been shown in \cite{loramath} that the $2^{SF}$ basis functions are orthogonal allowing a sufficiently synchronised receiver to demodulate using correlation. If $r_k$ is the received symbol corrupted by Additive White Gaussian Noise (AWGN), $\phi_i$, and $w^*_i$ is the complex conjugate of symbol $k$ (i.e. that corresponding to the transmitted symbol), the correlator output $y$ will exhibit a peak at index $k$.

\begin{equation}
    \sum_{n=0}^{2^{SF-1}} r_k(nT)  \cdot  w^*_i(nT) = \begin{cases}
      \sqrt{E_s} + \phi_i & i=k\\
      \phi_i & i \neq k\\
    \end{cases}     \label{awgncorr}  
    \end{equation}

\begin{equation}
y_k = \argmax (|\delta_{k,i} \sqrt{E_s} + \phi_i|) \label{awgncorr2}
    \end{equation}

It is demonstrated in \cite{loramath} how the more computationally efficient method removes the need to perform the full correlation over all $2^{SF}$ basis functions. Indeed, the method used by LoRa requires only the multiplication of $r_k$ with the complex conjugate of the base down chirp (a process known as dechirping). The dechirped signal comprises a pure frequency tone which is proportionate to $k$, so an FFT and find max routine completes the demodulation process.

Equation \ref{awgncorr} shows that correct demodulation will take place when $\sqrt{E_s} + \phi_k$ exceeds the maximum value of $\phi$ across all correlations. Although there may be significant distance between the PDFs of $\phi$ and $\sqrt{E_s} + \phi_k$, it is actually the PDF of the maximum of $\phi$ per symbol that is of interest in Symbol Error Rate (SER) calculations.

\subsection{Performance of LoRa in Multipath Channels}

\begin{figure}
\centering
\includegraphics[width=1\columnwidth]{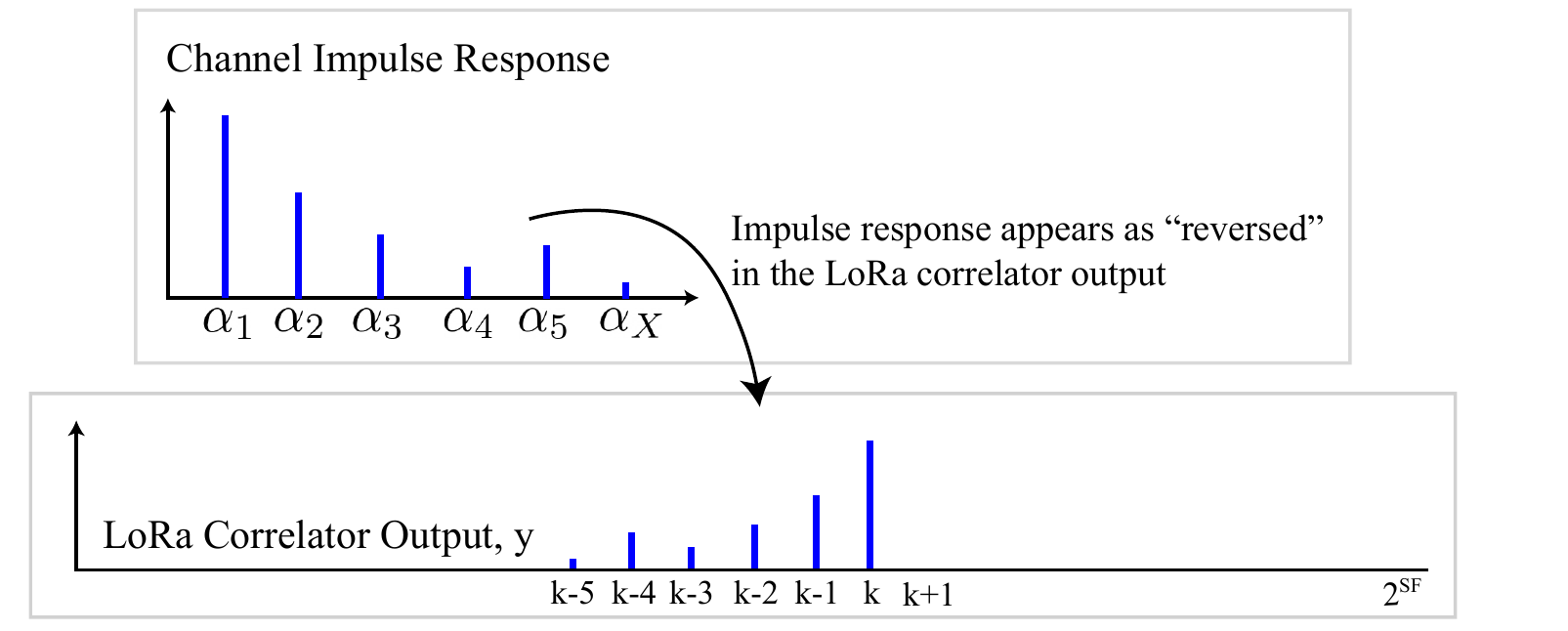} \\
\caption{The correlator output resulting from transmission in a multipath channel.}\label{pic}
\end{figure}

In most conventional LoRa applications, the RMS delay spread is in the ns or low $\mu$s range. This contains the majority of the multipath energy within a single sample and can be considered as frequency flat fading. However, in PLC applications, delay spreads of several tens of $\mu$s have been recorded. In this case, the multipath energy will smear across samples and will be more appropriately modelled as frequency selecting fading. A basic model of the situation can be constructed in which delayed versions of the transmitted symbol arrive at the receiver. Due to the unique way LoRa is modulated - effectively as time-shifted versions of a base chirp - a delayed version of a transmitted symbol will be demodulated as if it belonged to an adjacent symbol, as shown in Eqn.~\ref{multi_eqn}, where index $i=k-1 \dots k-x$ are proportionate to $\alpha_{2}\dots \alpha_x$, where $\alpha$ is the impulse response of the channel. This is shown graphically in Fig.~\ref{pic}.

\begin{equation}
    \sum_{n=0}^{2^{SF-1}} r_k(nT)  \cdot  w^*_i(nT) = \begin{cases}
      \sqrt{\alpha_1 E_s} + \phi_i & i=k\\
      \sqrt{\alpha_2 E_s} + \phi_i & i=k-1\\
		\vdots	& \vdots\\
			 \sqrt{\alpha_x E_s} + \phi_i & i=k-x\\
			  \phi_i & \text{elsewhere}\\
    \end{cases}     \label{multi_eqn}  
    \end{equation}
		
 Therefore, the channel impulse response can be  mapped out from the correlator output. 	In standard LoRa modulation, the presence of strong multipath interference in which the signal arrives by one or more indirect paths presents a problem for the LoRa demodulator because strong correlation peaks caused by these paths will compete against the true transmitted symbol, raising the SER.

\begin{figure*}
\centering
\includegraphics[scale=0.43]{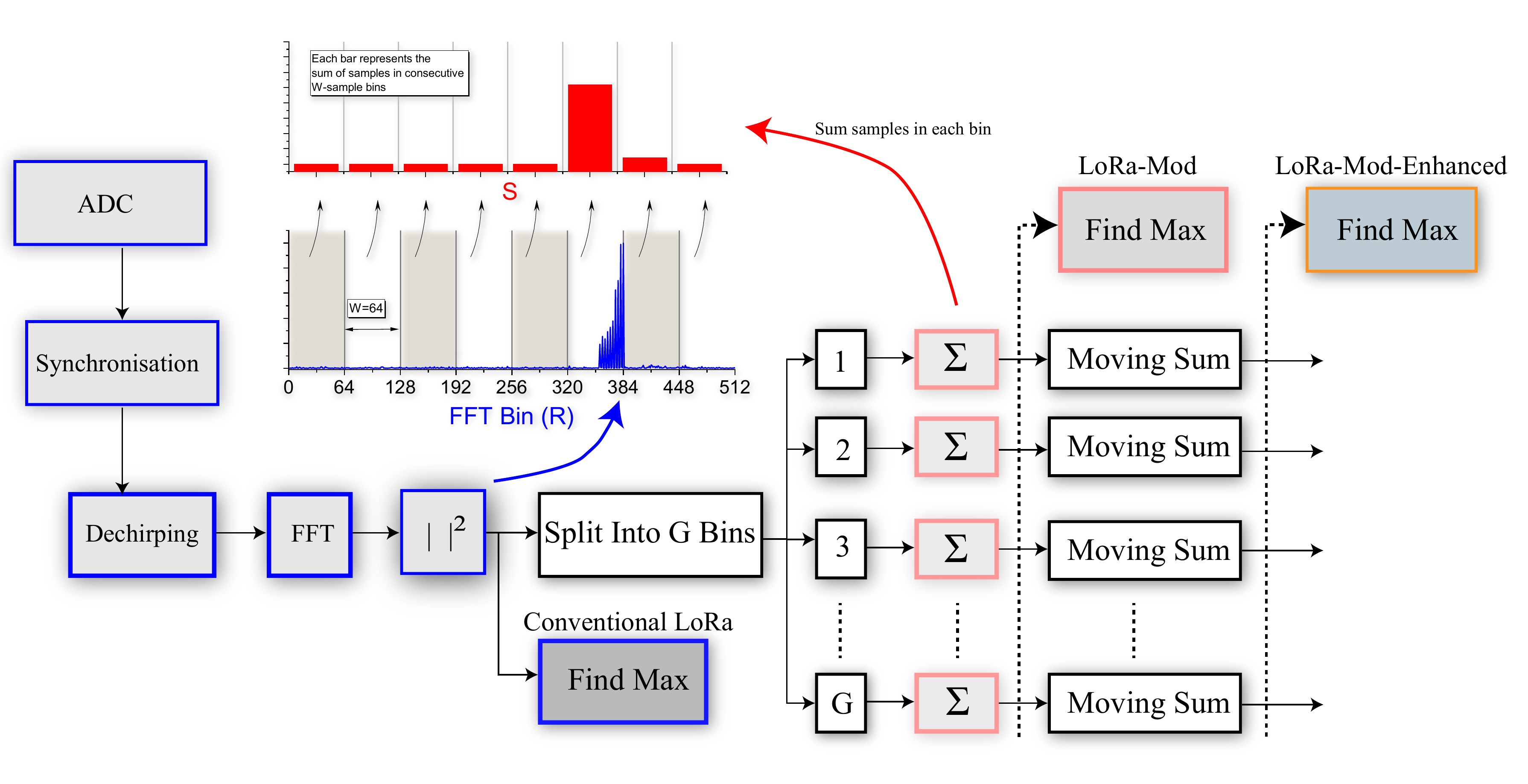} \\
\caption{Schematic diagram showing the proposed receiver architecture. The deviation from conventional LoRa starts at the ``Split into G Bins'' block. This subdivides the $2^{SF}$ bins into a reduced set of  $G = \frac{M}{P}$ `superbins'.  }\label{schem}
\end{figure*}

\section{Description of the Proposed Method}
\subsection{Enhancement for Robustness in Extreme Multipath}
It is interesting to note that the correlator output of the LoRa demodulator mimics the channel impulse response. This was noted in \cite{multipath_lora}, which exploits the regularity of the channel impulse response across subsequent LoRa symbols through the use of cross-correlation. This method would be particularly interesting in PLC applications because the channel is fixed. However, the performance is similar to that of conventional LoRa systems in the AWGN channel, which is still not good enough to perform reliably on the particularly hostile LV-MV channel. We also remark that the synchronisation requirements match that of conventional LoRa, which is difficult to achieve on the power line channel.

In the proposed method, which is shown graphically in Fig.~\ref{schem}, the convenient grouping of multipath energy into a predictable place in the correlator output is exploited in a different way. If the correlator output is termed $y(k)$, where each term in $y$ represents the absolute value of one of $2^{SF}$ output bins of the FFT operation, we can group the bins into a reduced set of $g$ `superbins'.

Assuming that the length of the channel impulse response is significantly smaller than the symbol time, $T_s$, the set of $M$ possible symbols can be reduced to a smaller set of $G = \frac{M}{P}$ possible symbols, each encoding ${SF} - \log_2{P}$ bits, where $P  2^\mathbb{Z}$  is the number of samples in each superbin. The modulator now encodes into a reduced set of $g \in 0 \dots G-1$ possible positions. If $P T_e$ is longer than the channel impulse response, the spread of energy resulting from multipath interference will be contained to a single superbin.
To implement this scheme at the transmitter, data should be encoded into one of a set of possible symbols described by $m^g \in P, 2P, \dots GP$. The restricted set of symbols are separated by $P$ samples, which, if longer than the manifestation of the channel impulse response within $R_m$, will contain the multipath energy to within the symbol being transmitted.

At the receiver, the process is identical to standard LoRa demodulation except that a reduced set of $G$ symbols are derived from the sum of the previous $P$ bins:

\begin{equation}
  S(g) =   \sum_{n=(g-1)P}^{gP}  |y (n)|^2 \label{sum}
\end{equation}

Equation~\ref{sum} shows a set of $S$ output terms, with each term representing the sum of $P$ correlator output terms from the FFT output, $y$. The summation combines the multipath energy within the symbol into a single number. The receiver can commence to finding the maximum index within $S$ in  the same way as the conventional LoRa demodulation process finds the maximum $y$. Equation~\ref{multi_eqn} shows that $y$ is made up of the square root of the symbol energy ($\sqrt{E_s}$) when $i=k$ (in the case of LoRa transmission in the AWGN channel) and a  dispersed share of the symbol energy when $k-x \geq i \leq k$ in the case of transmission on the multipath channel with a delay spread of $x$ samples. Every term in $y$ is also made up of a complex zero-mean Gaussian noise process, $\phi$.


 The find max routine must now choose from a reduced set of terms. The energy in each term is now made up of the sum of $P$ noise terms and, in the case of the correct symbol, the dispersed symbol energy. If the condition that $P T_e > \sigma_{rms}$, the majority of the symbol energy will fall within a single term. Therefore, the find max routine can still demodulate the symbol even in multipath channels, albeit at the expense of a reduced data rate. 

Each superbin comprises the sum of $P$ squared noise term samples, $|\phi|^2$, from the correlator output of Eqn.~\ref{multi_eqn}. Since $|\phi|$ is a Rayleigh distributed random variable, its square is distributed exponentially, and the sum of $P$ of these samples follows a Gamma distribution. We have approximated this as a Gaussian distribution with $\mu$ equal to the expected value of the sum of the noise terms and $\sigma$ equal to the sum of the noise variance, $ S_{\phi} \sim \mathcal{N}(P\mu,\,P\sigma^{2})$.

The superbin which corresponds to the correct symbol, $\mathcal{E}$, follows a Rician distribution with shape parameter $k_\beta = E_s/2\sigma^2 = E_s/N_0$, where $N_0$ is the single-sided noise power spectral density.

The demodulation process performs a find max routine which can be simplified as the comparison between $\mathcal{E}$ and the maximum of all other $g-1$ symbols. The distribution describing the maximum of $g-1$ normally distributed random variables is denoted $M_\phi$. The expected value of $M_\phi$ can be approximated as $\sqrt{2\log(g-1)}$ standard deviations greater than the expected value of $S_{\phi}$. A correct demodulation is achieved if $\mathcal{E} > M_\phi$.

\begin{figure}
\subfigure[SNR = -15 dB \label{d15}]{\includegraphics[width=0.475\columnwidth]{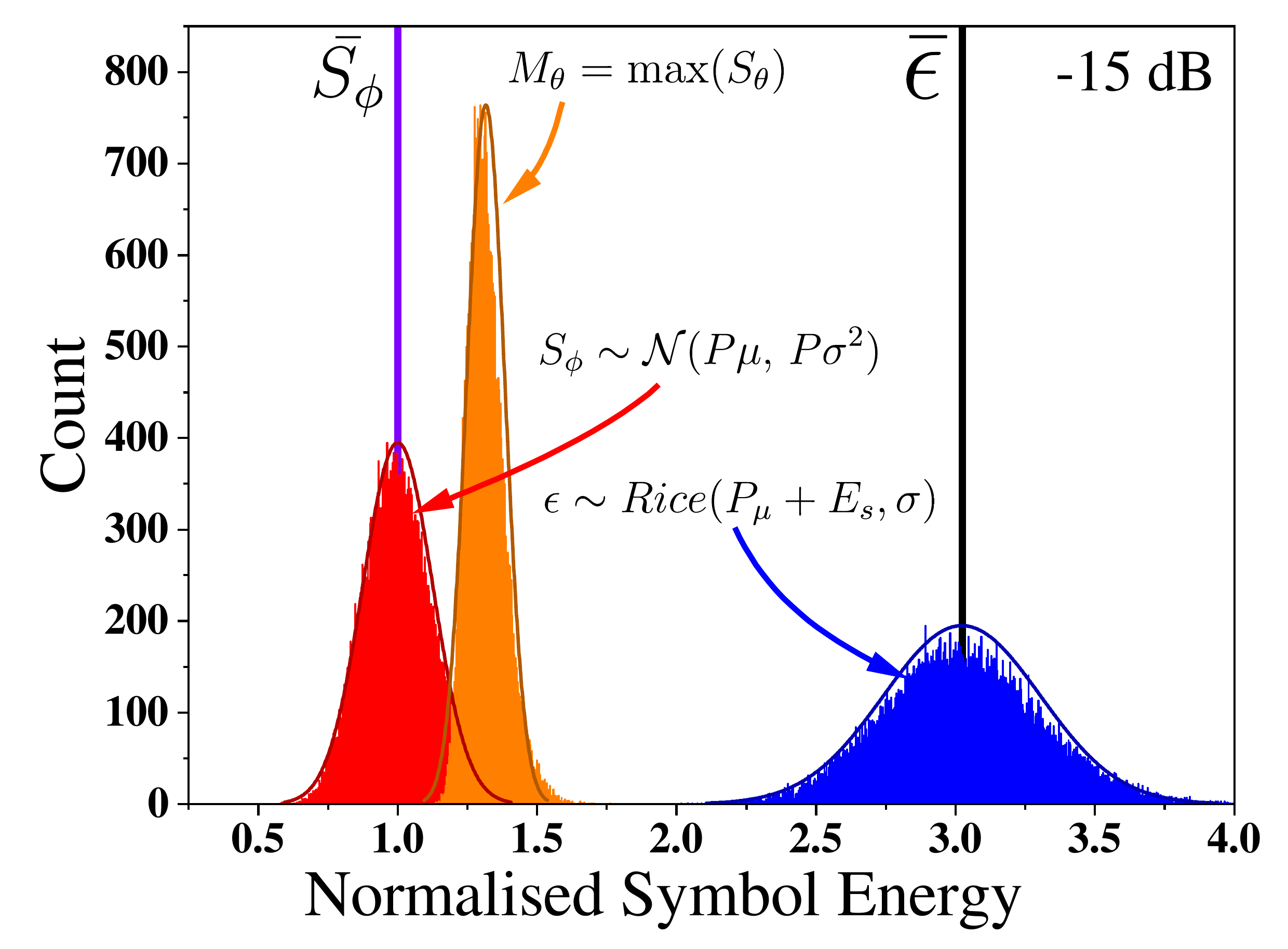}}
\subfigure[SNR = -20 dB \label{d20}]{\includegraphics[width=0.475\columnwidth]{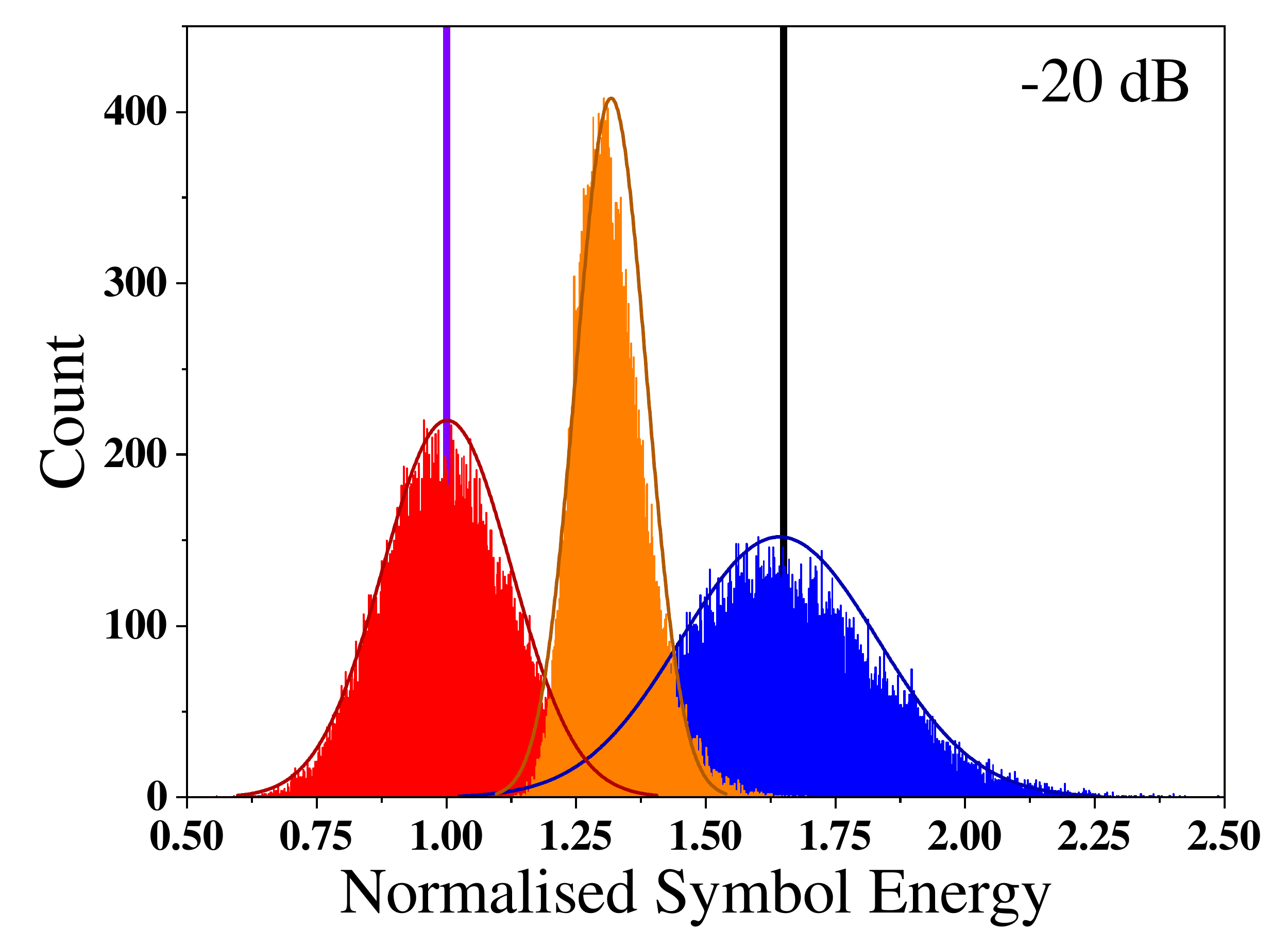}}
\\
\subfigure[SNR = -30 dB   \label{d30}]{\includegraphics[width=0.475\columnwidth]{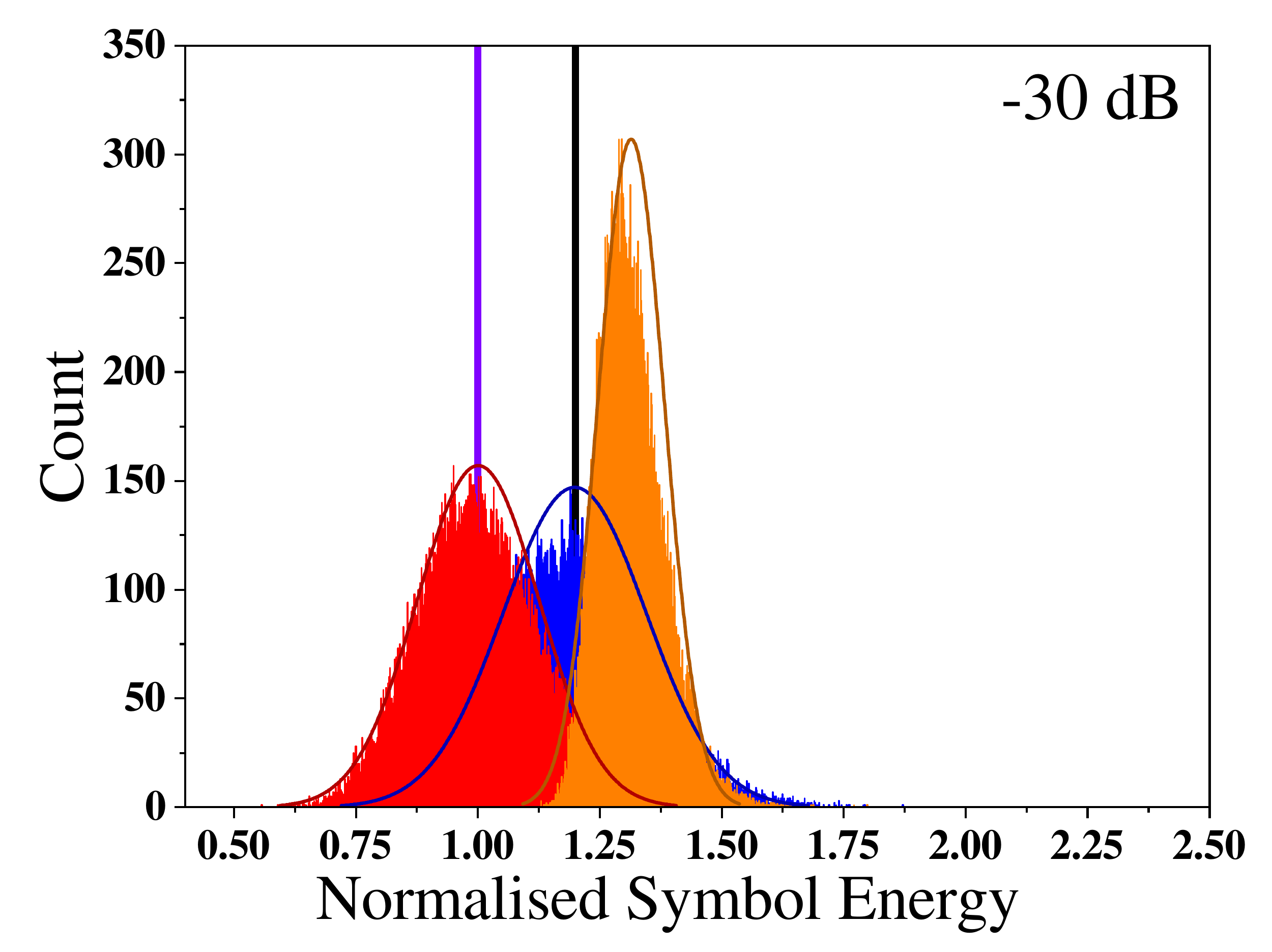}}
\subfigure[SNR = -35 dB   \label{d35}]{\includegraphics[width=0.475\columnwidth]{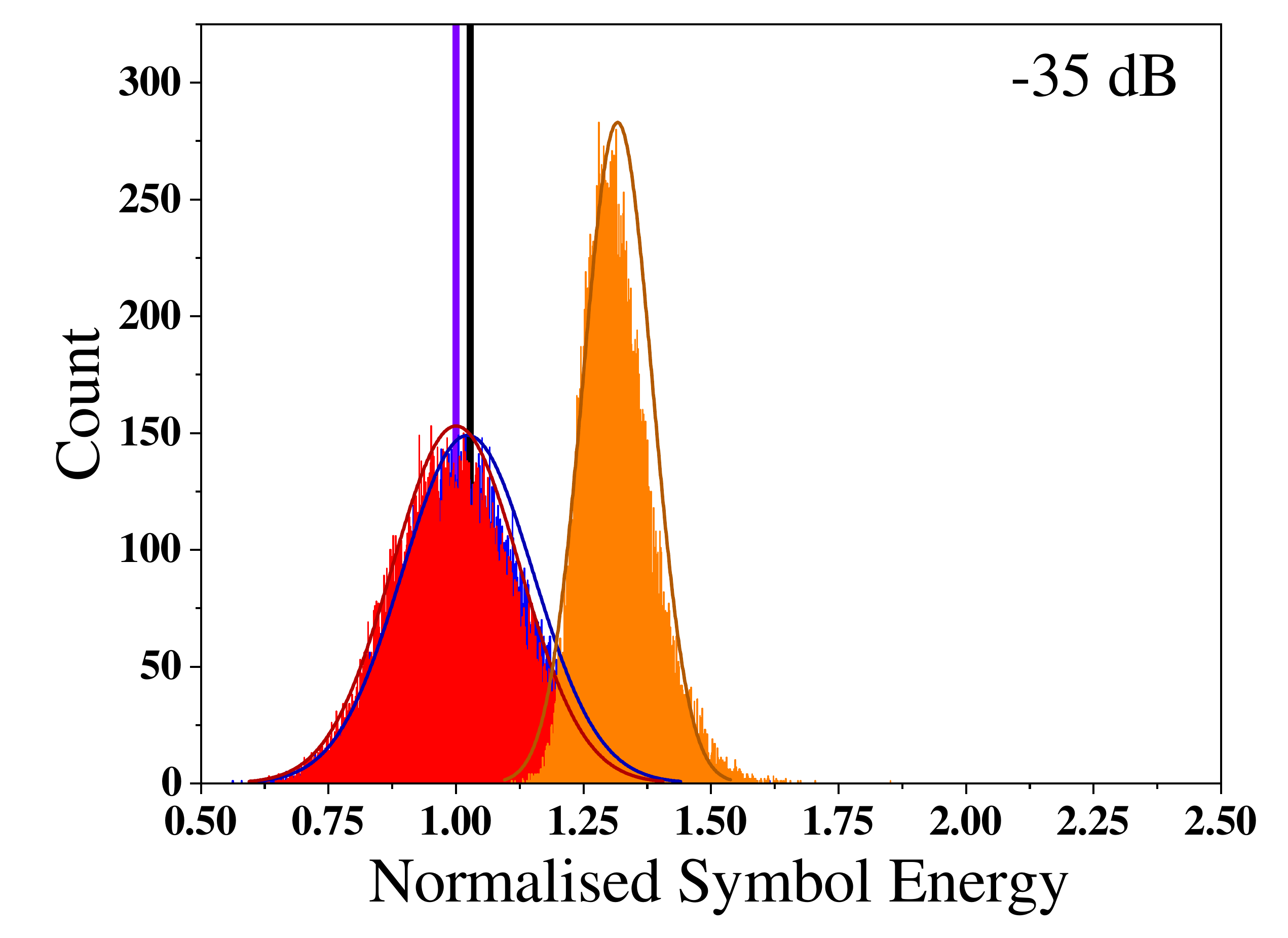}}
\\
\caption{Histograms Superbin size = 64, Quarter channel (worst case)}
\label{fig:histograms}
\end{figure}

Fig.~\ref{fig:histograms} shows $S_\phi$, $M_\phi$ and $\mathcal{E}$ resulting from 30,000 Monte Carlo simulations for a spreading factor of 12 and a superbin size of 64. A Rayleigh channel model with RMS delay spread $\tau$ = 100 $\mu$s is used, representing an equivalent bin size of 10 samples at an arbitrarily chosen 100 kHz LoRa sampling rate. This guarantees that most of the multipath energy will fall within a single superbin. The vertical reference lines labelled $\bar{S_\phi}$ and $\bar{\mathcal{E}}$ represent the means of the noise and transmitted symbol, respectively. The X scale is normalised to $\bar{S_\phi}$, and we retain this convention throughout the rest of the paper. 

 In Fig~\ref{d15} (SNR = -15dB), there is a clear separation between $S_\phi$ and $\mathcal{E}$. LoRa-Mod, which relies on a comparison between $\mathcal{E}$ and $M_\phi$,  can operate successfully at this SNR. Fig~\ref{d20} (SNR = -20dB) shows a more pronounced overlap between $\mathcal{E}$ and $M_\phi$, which leads to an increasing symbol error rate for LoRa-Mod. Beyond this point, LoRa-Mod cannot be used. However, it is noted that the mean of $\mathcal{E}$ is still larger than the mean of $S_\phi$, even at extremely low SNRs (e.g. in the case of -35dB in Fig.~\ref{d35}).

\begin{figure}
\subfigure[10 sample channel SF=12 \label{AA}]{\includegraphics[width=0.465\columnwidth]{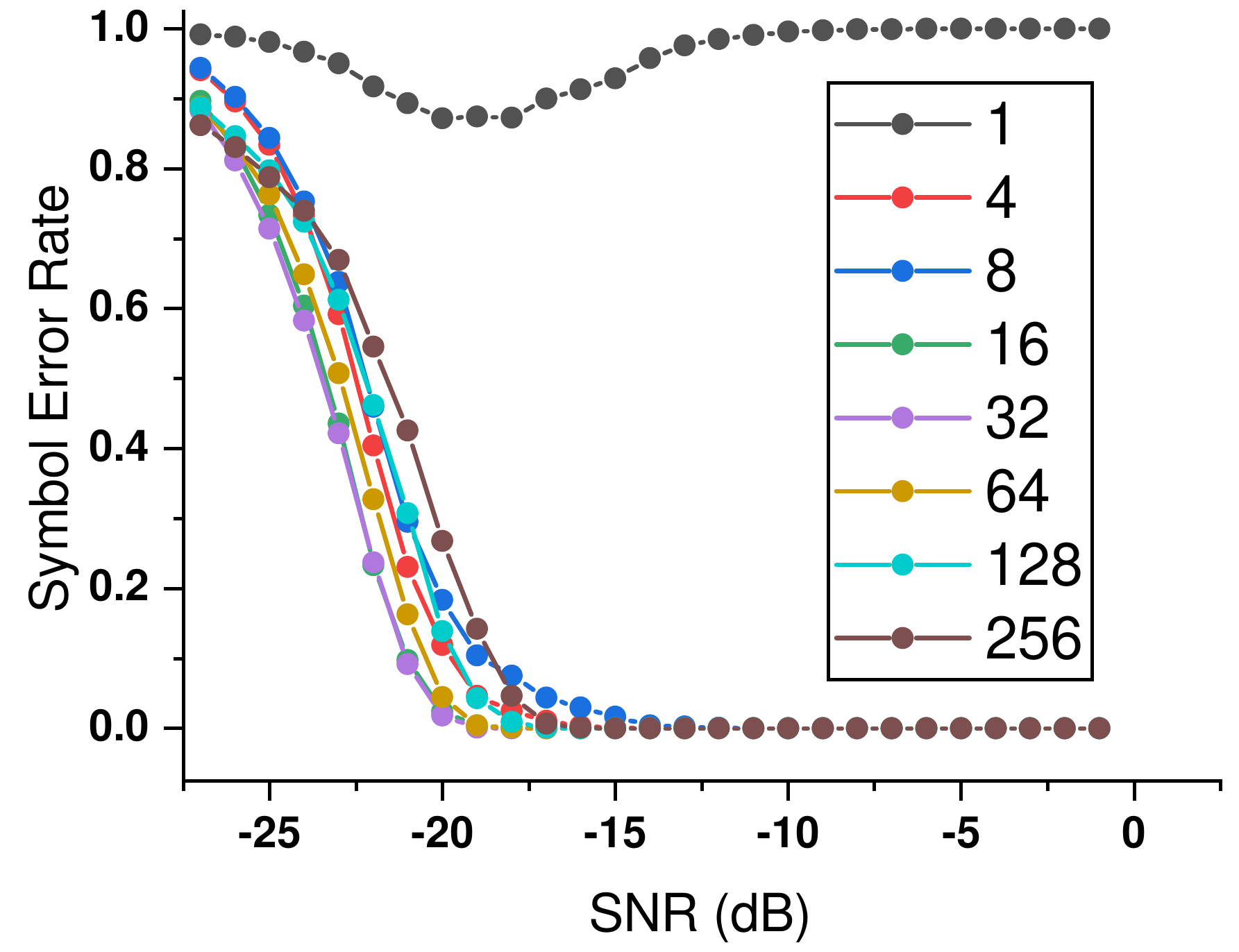}}
\subfigure[10 sample channel SF=14 \label{BB}]{\includegraphics[width=0.465\columnwidth]{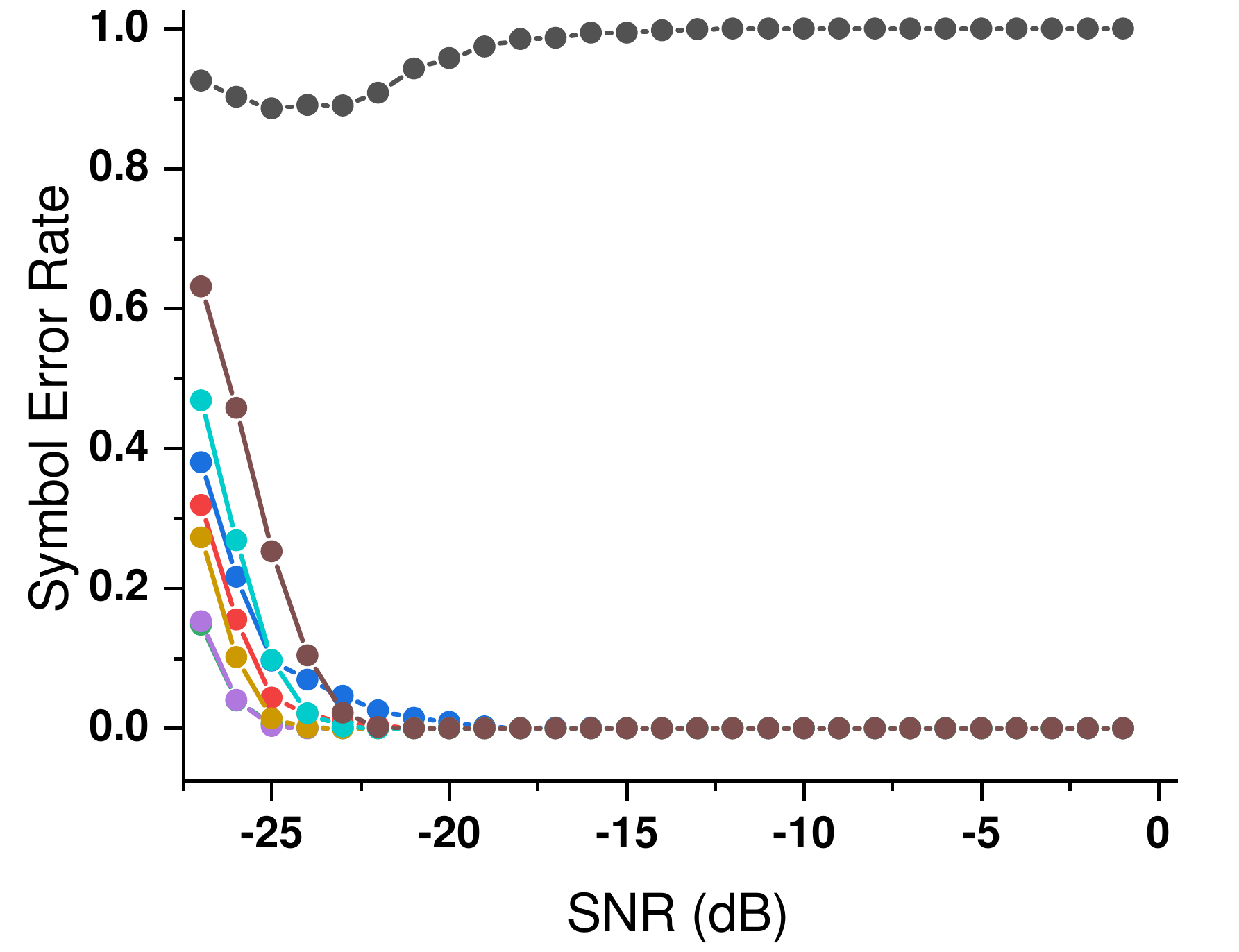}}
\\
\subfigure[20 sample channel SF=12 \label{CC}]{\includegraphics[width=0.465\columnwidth]{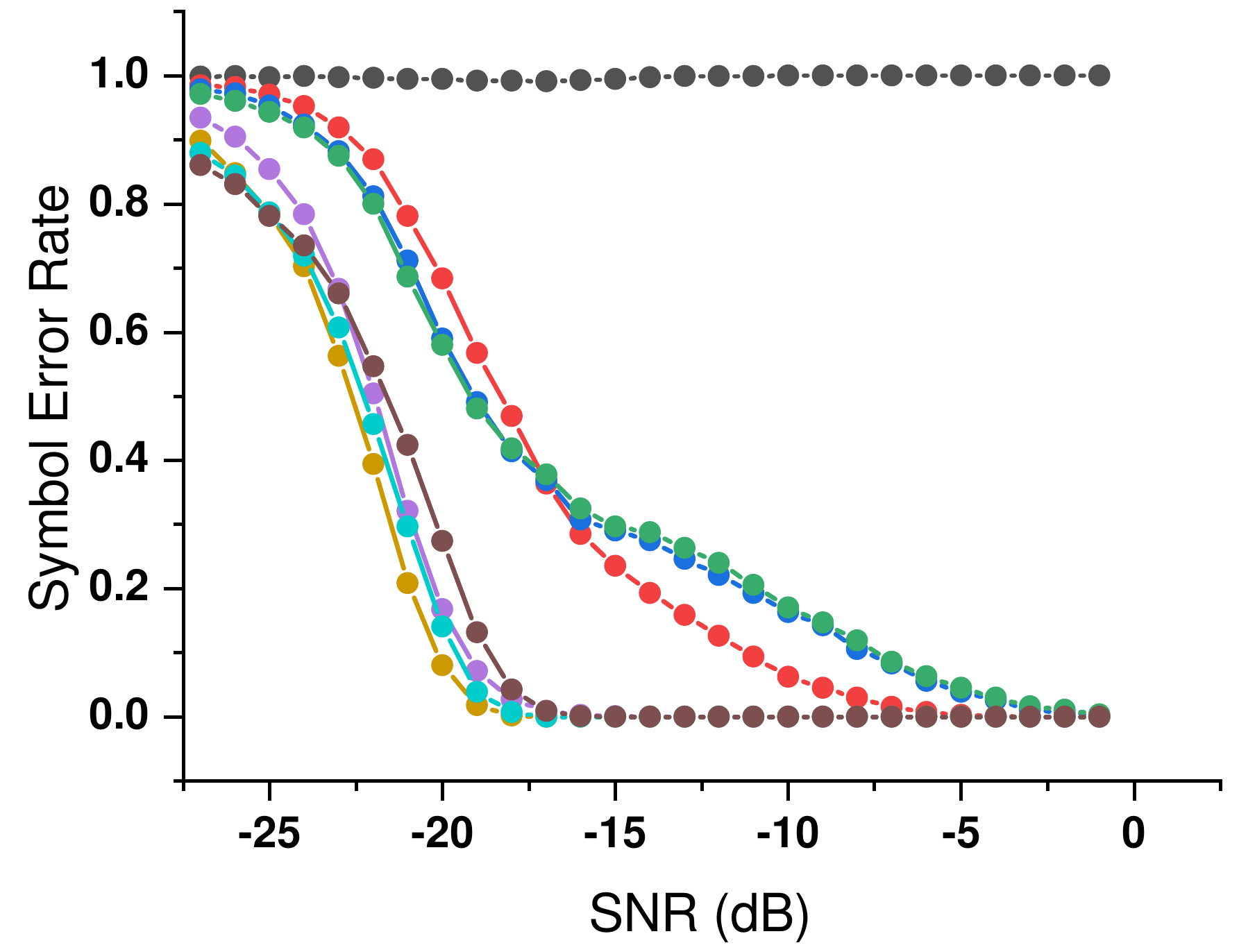}}
\subfigure[20 sample channel SF=14  \label{DD}]{\includegraphics[width=0.465\columnwidth]{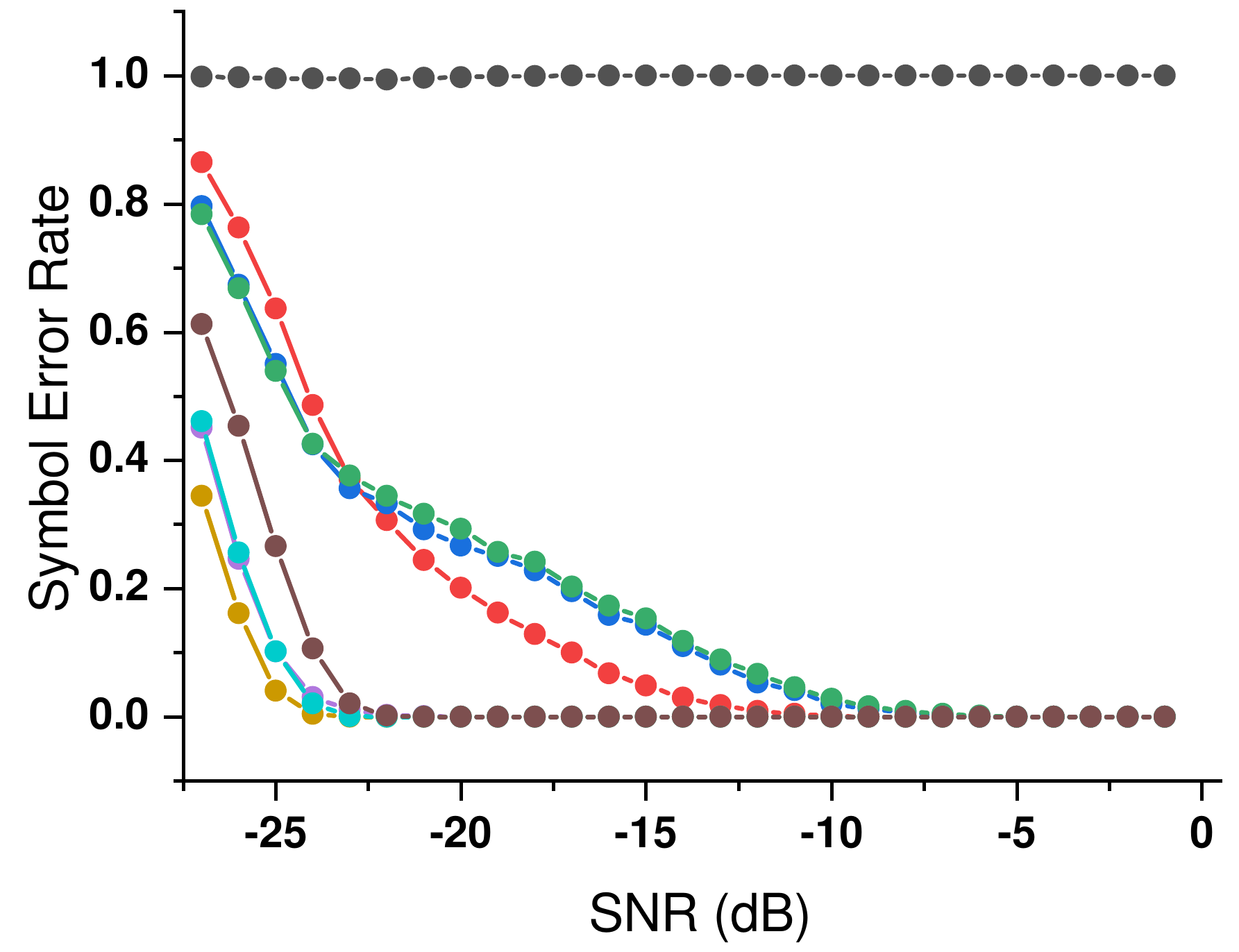}}
\\
\subfigure[40 sample channel SF=12 \label{EE}]{\includegraphics[width=0.465\columnwidth]{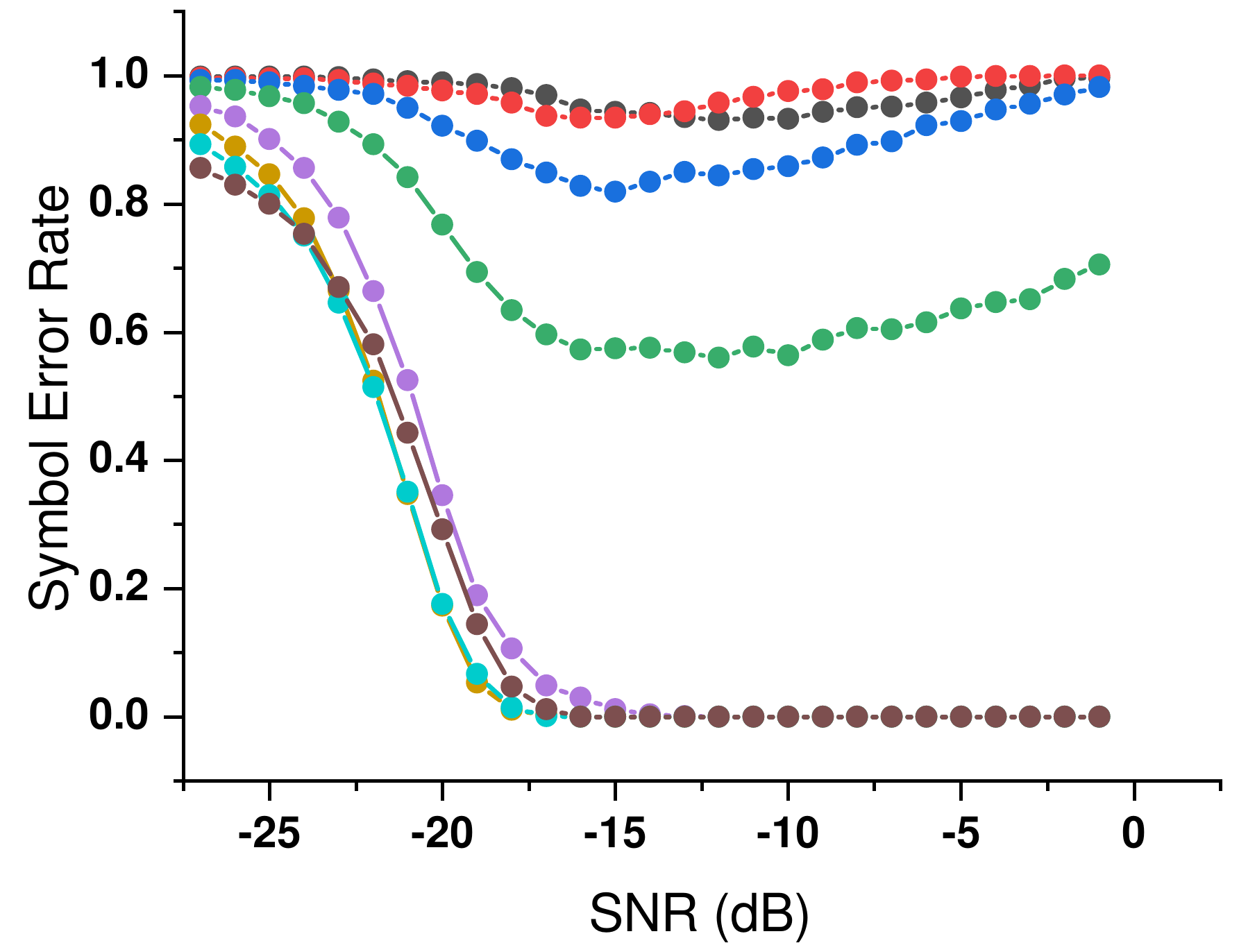}}
\subfigure[40 sample channel SF=14  \label{FF}]{\includegraphics[width=0.465\columnwidth]{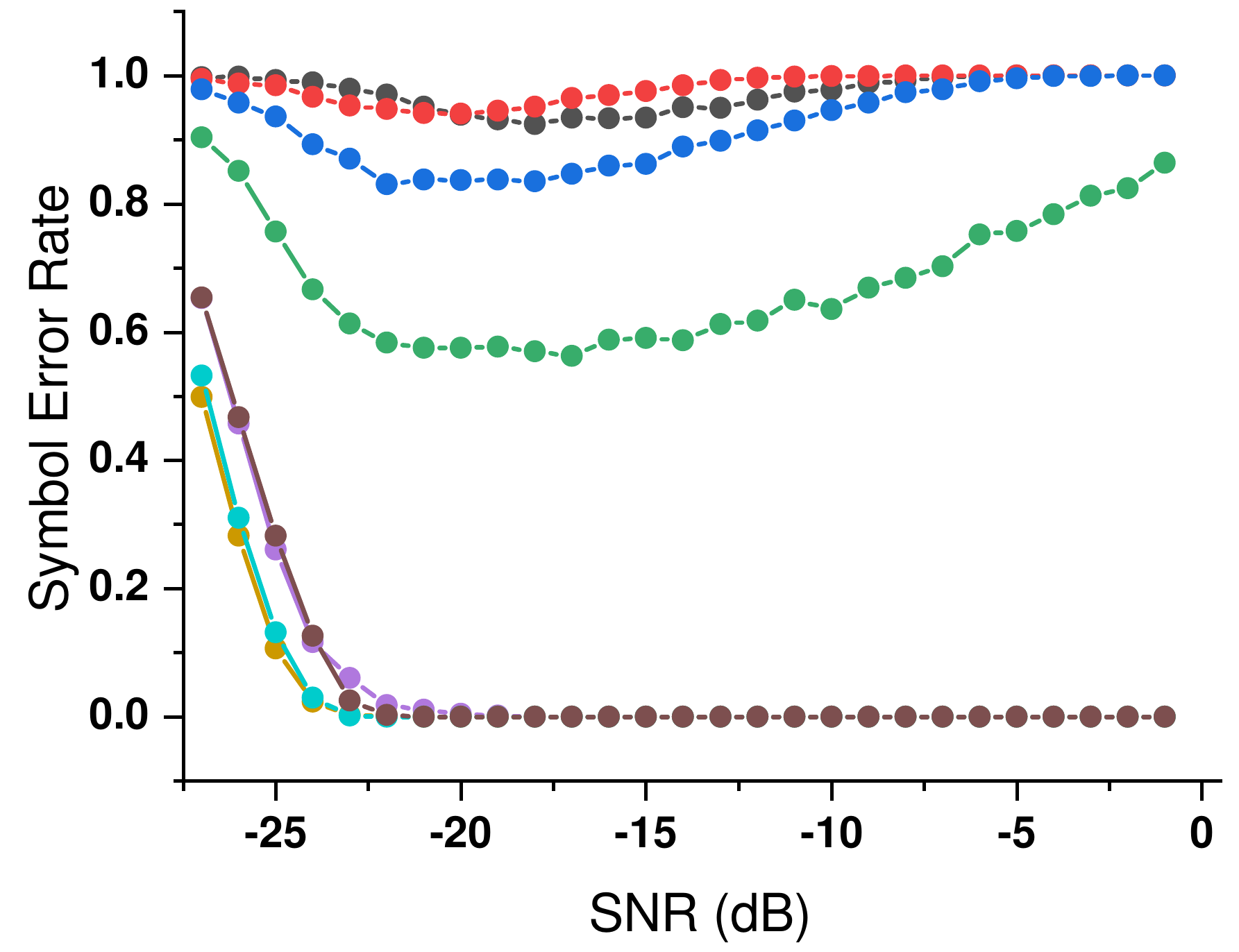}}
\caption{\textbf{Performance of LoRa-Mod:} SNR Versus Symbol Error Rate as a function of superbin size, SF=12, 14, and various length Rayleigh channels.}
\label{fig:LORAMODPERF}
\end{figure}

Fig.~\ref{fig:LORAMODPERF} shows the performance of LoRa-Mod in three separate Rayleigh channels ($\tau_{rms}$ = 10, 20 and 40 LoRa samples) and for two different spreading factors (SF=12, 14). Firstly, the general level of performance improves with increasing spreading factor. Secondly, there is a requirement for the superbin size to be at least as large as $\tau_{rms}$. For example, in Fig.~\ref{AA} and \ref{BB}, which is the 10 sample channel, all superbin sizes perform relatively well. In the 20 sample channel (Fig.~\ref{BB} and \ref{DD}), the 4, 8 and 16 sample binsize schemes exhibit a drop in performance. In the 40 sample channel, the trend continues.

\subsection{Enhancement for Robustness in Extremely Low SNRs}
Lora-Mod has the flexibility to perform well in channels with arbitrarily long RMS delay spreads and poor SNRs, but a further trade-off of data rate can improve the performance to even lower SNR regimes. This scheme, termed Lora-Mod-Enhanced henceforth, uses the principle of statistical averaging to estimate the means of $\mathcal{E}$ and the noise symbol $S_\phi$ distributions. As shown in Fig.~\ref{schem}, this comes at the cost in hardware of a recursive running sum or running mean for each superbin.

In extremely low SNRs, the difference between the means of $\mathcal{E}$ and $S_\phi$ is small but non-zero. Should the same transmitted symbol be repeatedly sent, the receiver can estimate the mean of the previous $Q$ symbols of $\mathcal{E}$ (the transmitted symbol) and $S_\phi$ (all noise symbols). Unlike LoRa-Mod, which is a comparison between the maximum noise symbol ($M_\phi$, shown in orange in Fig. \ref{fig:histograms}), and $\mathcal{E}$, LoRa-Mod-Enhanced is a comparison between the mean of the previous Q $\mathcal{E}$ symbols with the maximum of the means of the previous Q $S_\phi$ symbols. 

\begin{equation}
\mathcal{H}(g) = \frac{1}{Q} \sum_{n=0}^{Q} S(g-n)
\end{equation}

The variance of $\mathcal{H}$ is narrowed compared to $S_\phi$:

\begin{equation}
 \bar{S_\phi} = H_{\phi} \sim \mathcal{N} \left ( P\mu,\,\frac{P\sigma^{2}}{Q} \right )
\end{equation}

Although the symbol error rate is still dependent on the comparison between the transmitted symbol and the maximum of the noise symbols, both terms have distributions which are significantly narrowed by the averaging.

Since it can be guaranteed that most of the transmitted symbol energy will fall within $\mathcal{E}$, a high enough Q is able to detect the statistical difference between the two means, even at extremely low SNRs.

This approach requires a compromise in terms of the maximum achievable datarate since, in effect, any gains made in robustness due to an increase in Q are matched by a proportionate reduction in datarate. However, in grid applications where load data is only required on the timescale of minutes, this comprimise might be acceptable.

\begin{figure}
\centering
\includegraphics[width=1\columnwidth]{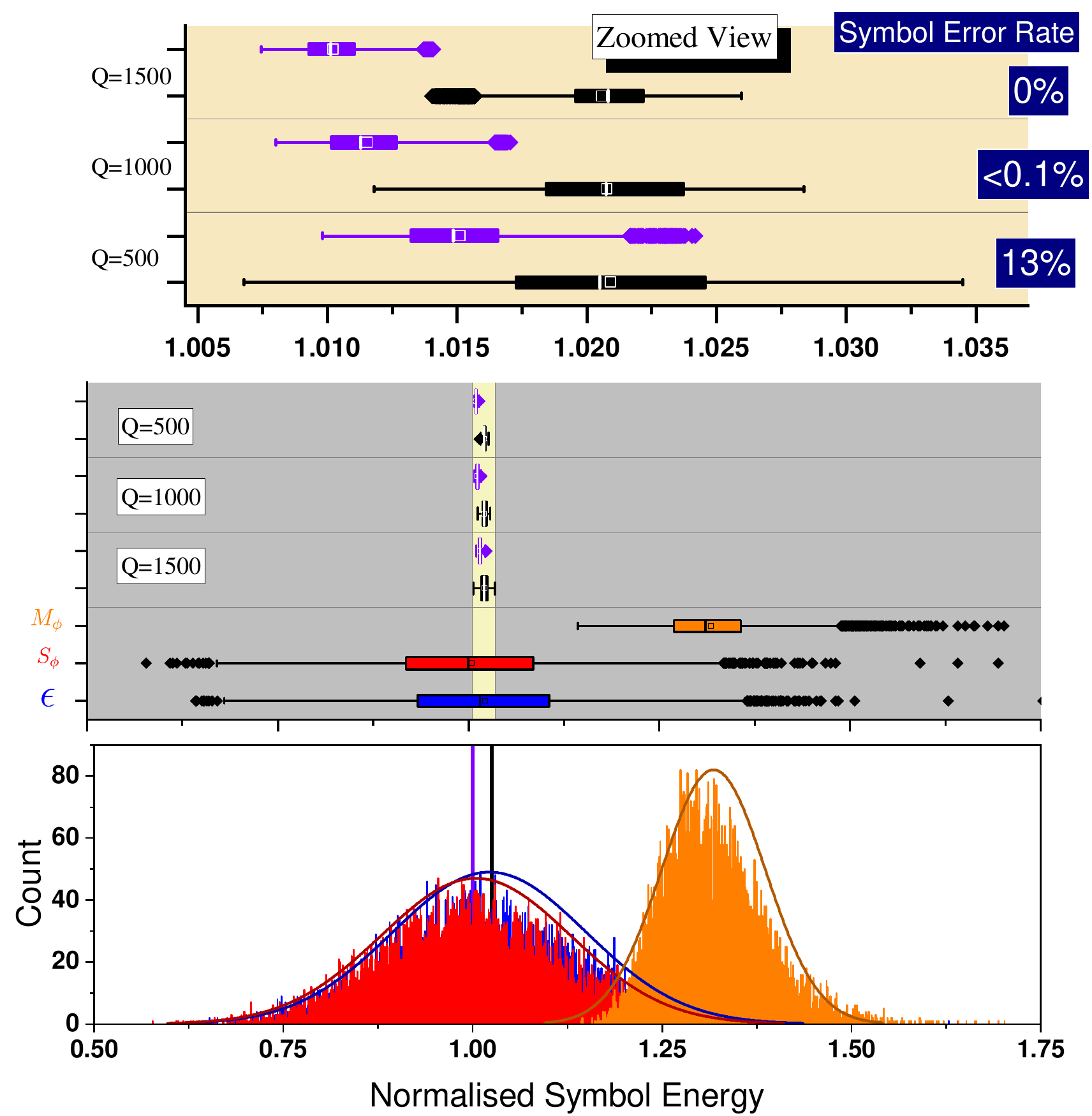} \\
\caption{\textbf{Demonstration of how LoRa-Mod-Enhanced can provide robust communication in a low SNR environment (-40 dB here)} .The bottom panel shows the distribution of $S_\phi$ (red), $\mathcal{E}$ (blue) and $M_\phi$ (orange). The middle panel shows the boxplot version of these distributions, alongside $\mathcal{H}$ for Q=1500, 1000 and 500. The top panel is a zoomed view of $\mathcal{H}$, showing clearly how the symbol and noise distributions separate with increasing Q. These simulations are performed with the Rayleigh 20 sample channel, SF=12}\label{boxnotm}
\end{figure}

\begin{figure}
\centering
\includegraphics[width=0.85\columnwidth]{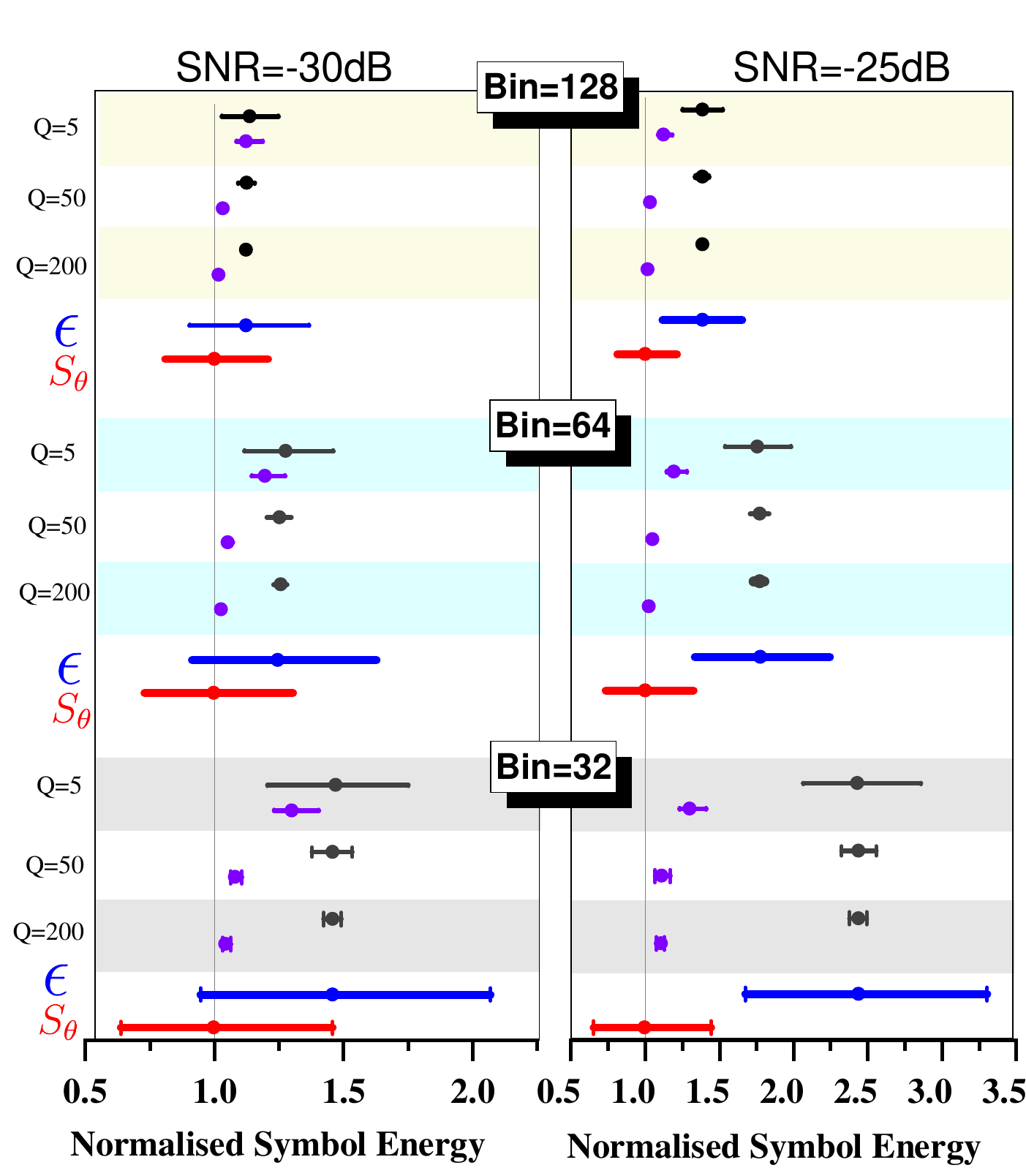} \\
\caption{Boxplots summarising the distributions of $S_{\phi}$, $\mathcal{E}$ and $\mathcal{H}$ (LoRa-Mod-Enhanced) for Q=5, 50 and 500 symbols. Increasing Q narrows the distribution of $\mathcal{H}$ and creates a wider normalised separation, making error free communication possible.}\label{boxnotm2}
\end{figure}

Fig.~\ref{boxnotm2}, which retains the colour scheme from Fig.~\ref{boxnotm},  shows the distributions (represented as horizontal lines) of normalised symbol energy for a spreading factor of 12 and for various bin sizes. It also shows the effect of varying the running mean length, Q. The same LoRa parameters deployed in Fig.~\ref{boxnotm} are used, meaning the majority of the multipath energy falls within the transmitted symbol superbin, $\mathcal{E}$. It is interesting to note that increasing the bin size beyond that which is necessary to contain the multipath energy actually deteriorates performance of LoRa-mod-enhanced, as indicated by the narrowing of the gap between the purple lines ($\bar{S_\phi}$ and the $\bar{\mathcal{E}}$).


\subsection{Synchronisation}
The synchronisation requirements are drastically relaxed in comparison to conventional LoRa because it does not matter if energy smears across LoRa bins (the energy will remain inside the superbin). This opens up the possibility of using zero-crossing detectors to estimate the start point of each symbol, reducing the complexity of both transmitter and receiver and removing the requirement for a preamble.

\section{Case Study}
\subsection{Development of a test network}
\begin{figure*}
\centering
\includegraphics[width=0.99\textwidth]{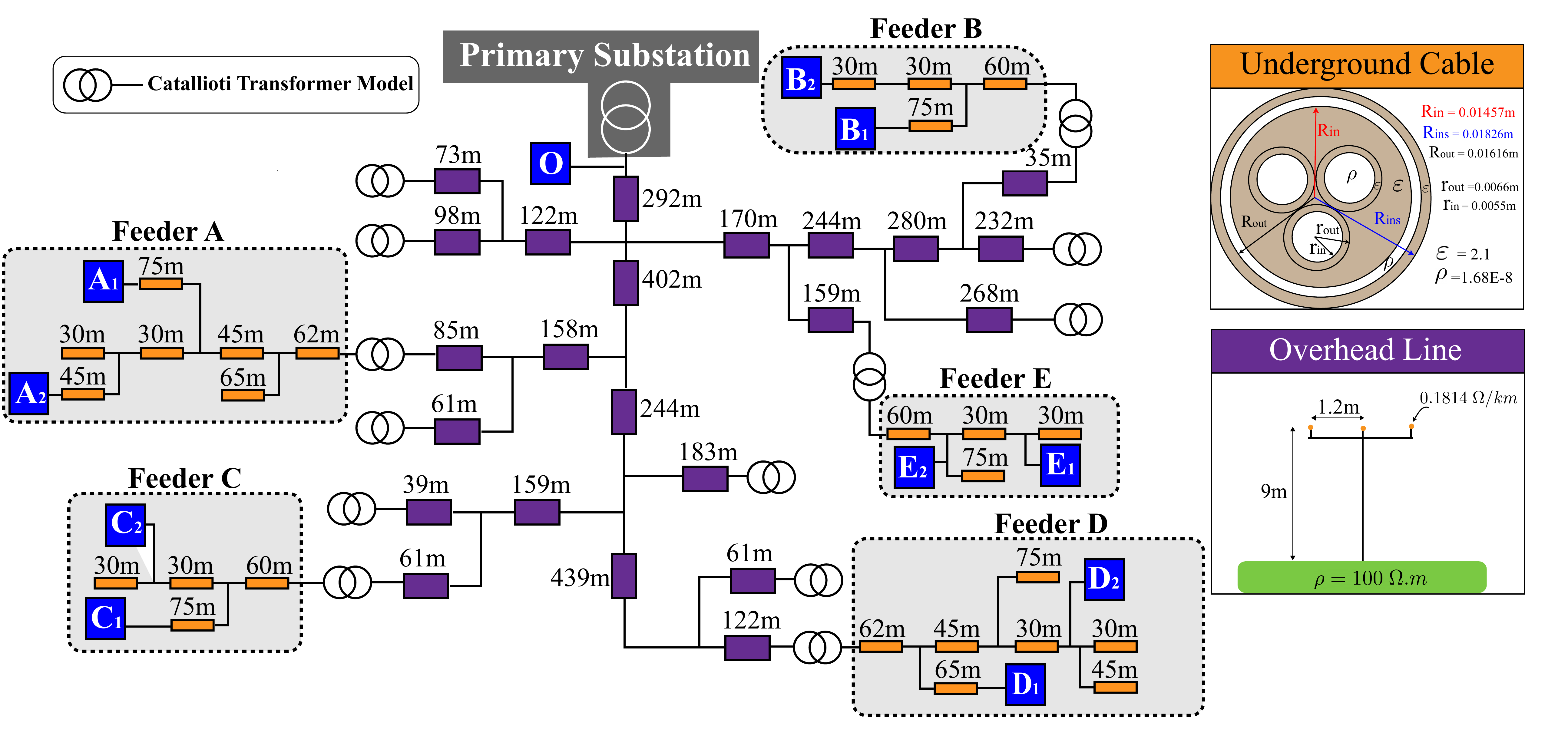} \\
\caption{ATP-EMTP Test Network comprising 5 LV feeders and a radial MV network.}\label{TESTNETWORK}
\end{figure*}

The test network, as shown in Fig.~\ref{TESTNETWORK}, has 5 LV feeders and a radial MV network. The LV feeders use underground cable models (three-phase in an enclosed pipe) and the HV network comprises an overhead line based on an 11 kV wood pole model. All models use the frequency dependent JMarti model generated by the Lines and Cables Constants (LCC) routine of the EMTP. The MV/LV transformers are modelled with the high frequency Catallioti model \cite{tx_model} and each LV line is terminated on the load side by a three-phase 10 $\Omega$ resistance.

\subsection{Simulation Methodology}
Two LoRa-Mod-Enhanced devices are placed within each of the 5 feeders (labelled A-E in Fig.~\ref{TESTNETWORK}). A single receiver is placed near the primary substation (labelled `O'). The modulation code, which is executed in Matlab, automatically writes each sample to a text file. This text file is read by a `foreign model', which is a custom piece of code executed natively within the EMTP, and provides the source signal for a TACS source. The process can be repeated for any number of TACS sources within the simulation, providing scope for a full system-wide simulation incorporating several transmitters. This methodology also resolves the challenge of writing many millions of samples into EMTP, and allows the communication signal to be read at the same $\Delta$T as the simulation (1E-7). On completion, the EMTP simulation results are converted to a .MAT file using the PL42MAT routine and processed by the demodulation code within Matlab. 

\begin{figure}
\centering
\includegraphics[width=0.76\columnwidth]{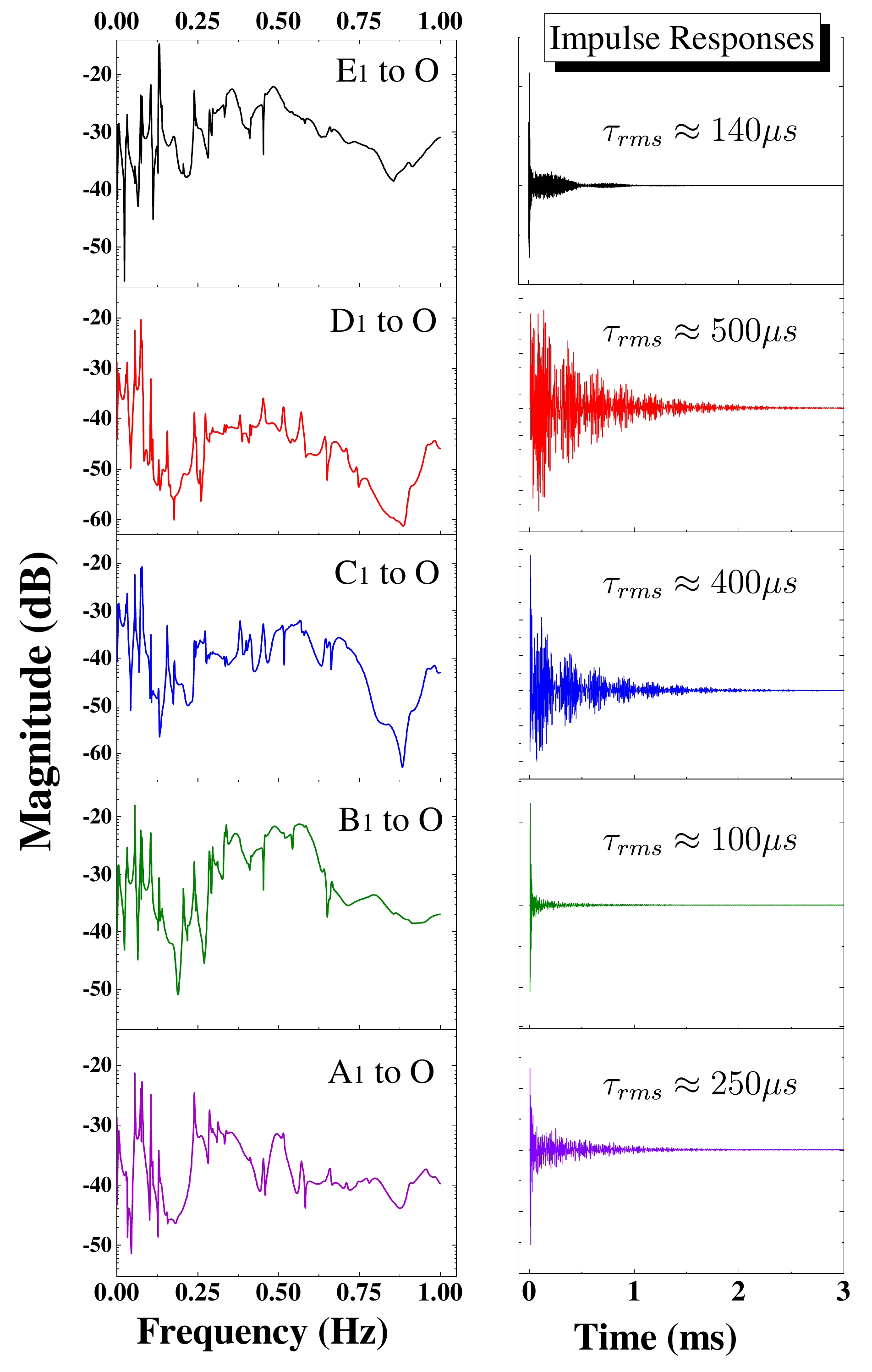} \\
\caption{Magnitude and Impulse responses for 5 selected channels separating feeders A-E from the observation point, O.}\label{imp_reso}
\end{figure}

Fig.~\ref{imp_reso} shows the magnitude and impulse responses between a selection of 5 transmitters and the observation point (`O'). The shape of these responses is representative of the MV power line channel, which is characterised by extreme multipath and regions of high attenuation. The RMS delay spread is observed to vary between 100 $\mu$s and 500 $\mu$s. In the context of a network-wide implementation of the proposed scheme, the magnitude responses show the importance of bandwidth selection. For example, fluctuations of just a few kHz exhibit differences of tens of dBs.

\begin{table}
\caption{Simulation Parameters \label{hhhh}}
\begin{tabular}{ccccccc}\toprule
 \multicolumn{2}{c}{} &  \multicolumn{3}{c}{Time on Air (s)} &  \multicolumn{2}{c}{} \\
\cmidrule(l){3-5}  
   Tx  & SF & Q=1 & Q=10 & Q=100 & $f_c$  & BW 	\\
	\midrule
$A_1$ & 13 & 0.33 & 3.3 & 33 &  50 kHz & 25 kHz\\
$A_2$ & 13 & 0.33 & 3.3 & 33 & 80 kHz & 25 kHz \\
$B_1$ & 13 & 0.33 & 3.3 & 33 & 110 kHz & 25 kHz \\
$B_2$ & 13 & 0.33 & 3.3 & 33 & 140 kHz & 25 kHz  \\
$C_1$ & 13 & 0.33 & 3.3 & 33 & 170 kHz & 25 kHz \\
$C_2$ & 13 & 0.33 & 3.3 & 33 & 210 kHz & 25 kHz \\
$D_1$ & 13 & 0.33 & 3.3 & 33 & 240 kHz & 25 kHz  \\
$D_2$ & 13 & 0.33 & 3.3 & 33 & 270 kHz & 25 kHz \\
$E_1$ & 13 & 0.33 & 3.3 & 33 & 300 kHz & 25 kHz \\
$E_2$ & 13 & 0.33 & 3.3 & 33 & 330 kHz & 25 kHz \\
\bottomrule
\end{tabular}
\end{table}

\subsection{Simulation Results}

\begin{figure*}[!b]
\subfigure[Transmitter \label{TX}]{\includegraphics[width=0.28\paperwidth]{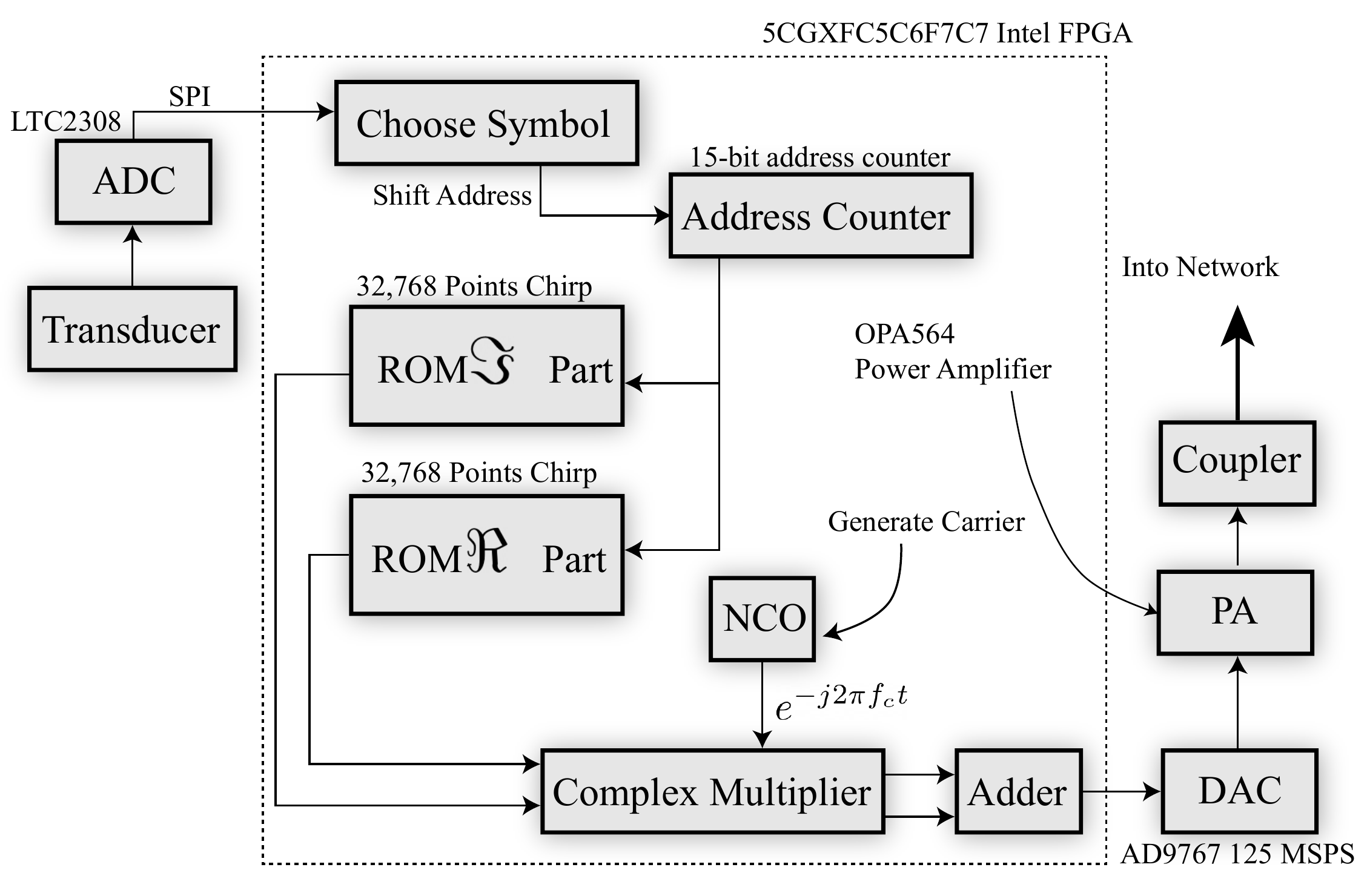}}
\subfigure[Receiver\label{RX}]{\includegraphics[width=0.55\paperwidth]{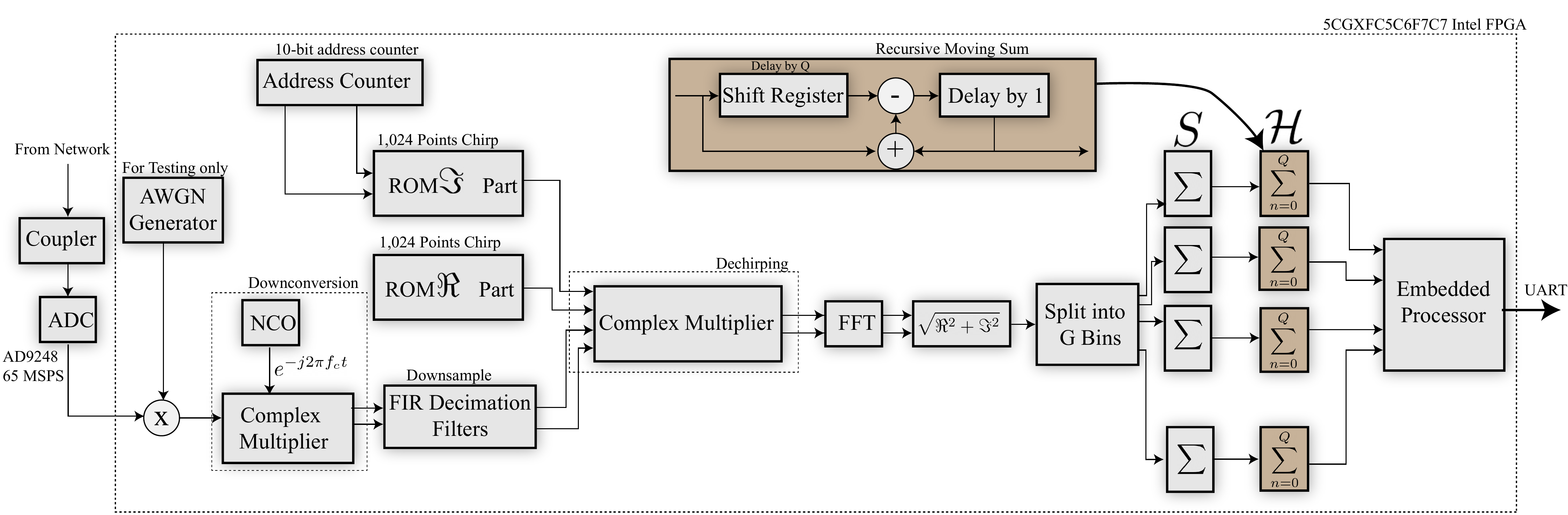}}
\caption{FPGA based hardware architectures for the transmitter and receiver.}
\label{fpga}
\end{figure*}

\begin{figure}
\subfigure[A1-O \label{SUMa}]{\includegraphics[width=0.327\columnwidth]{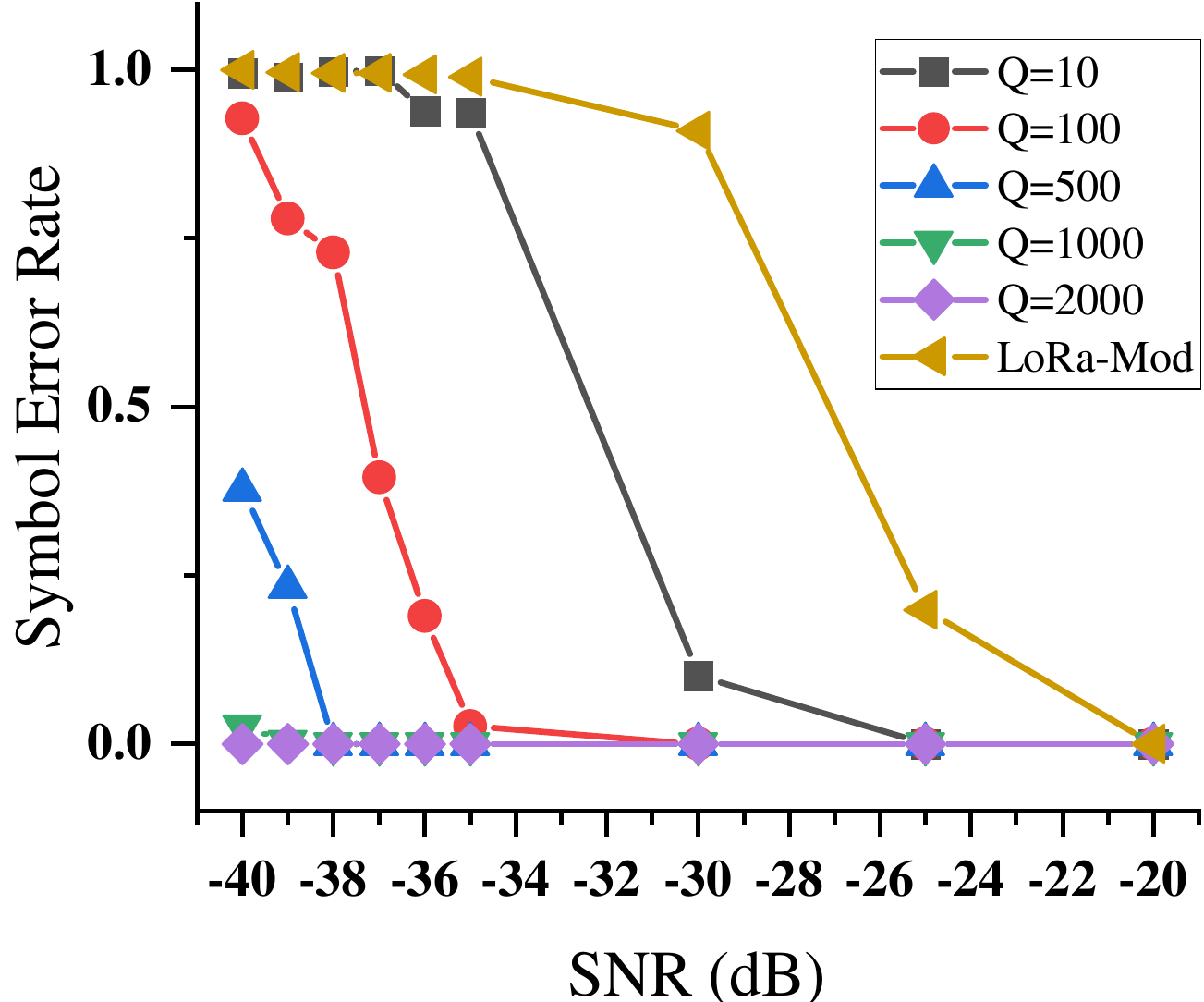}}
\subfigure[A2-O \label{SUMb}]{\includegraphics[width=0.327\columnwidth]{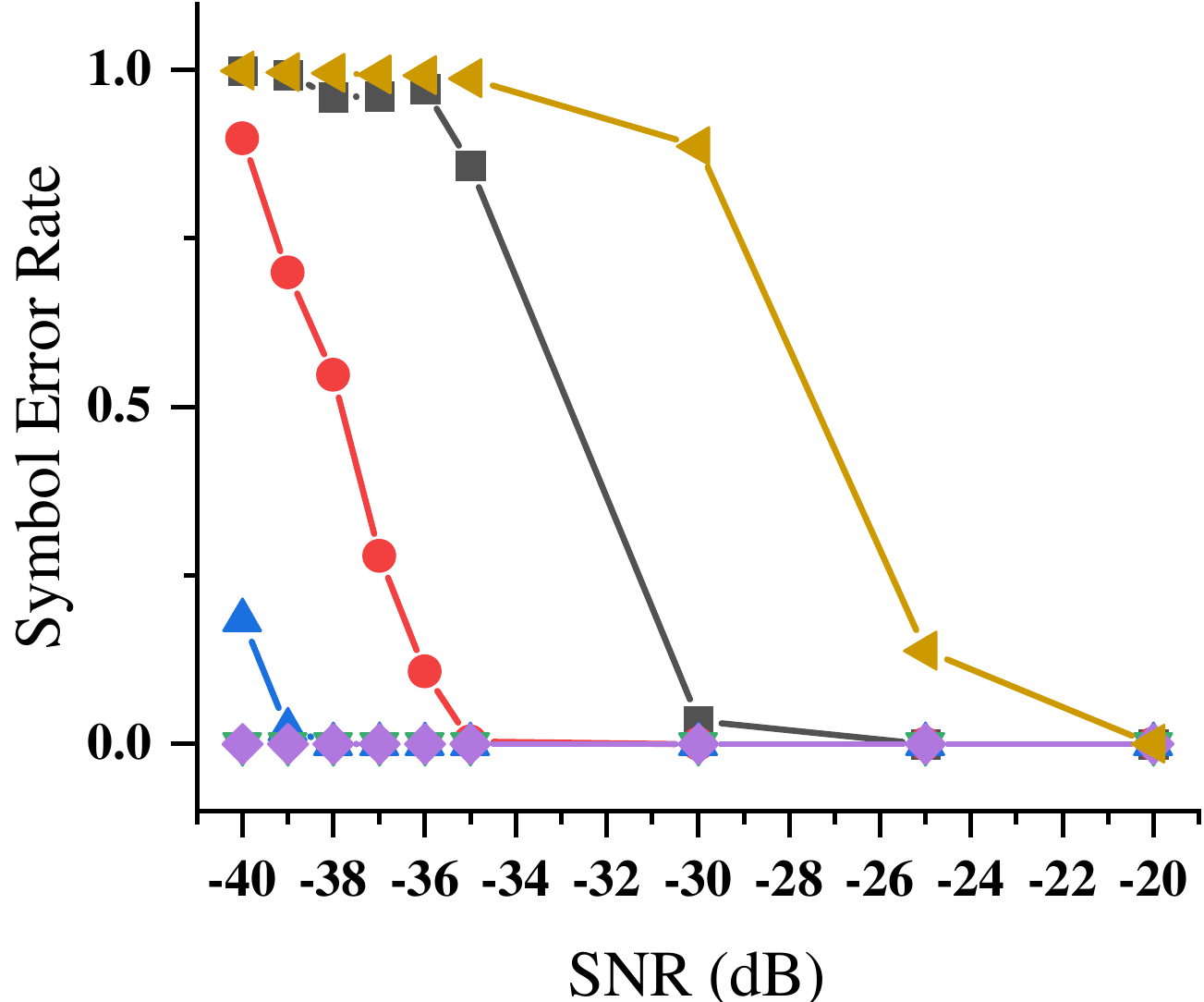}}
\subfigure[B1-O \label{SUMc}]{\includegraphics[width=0.327\columnwidth]{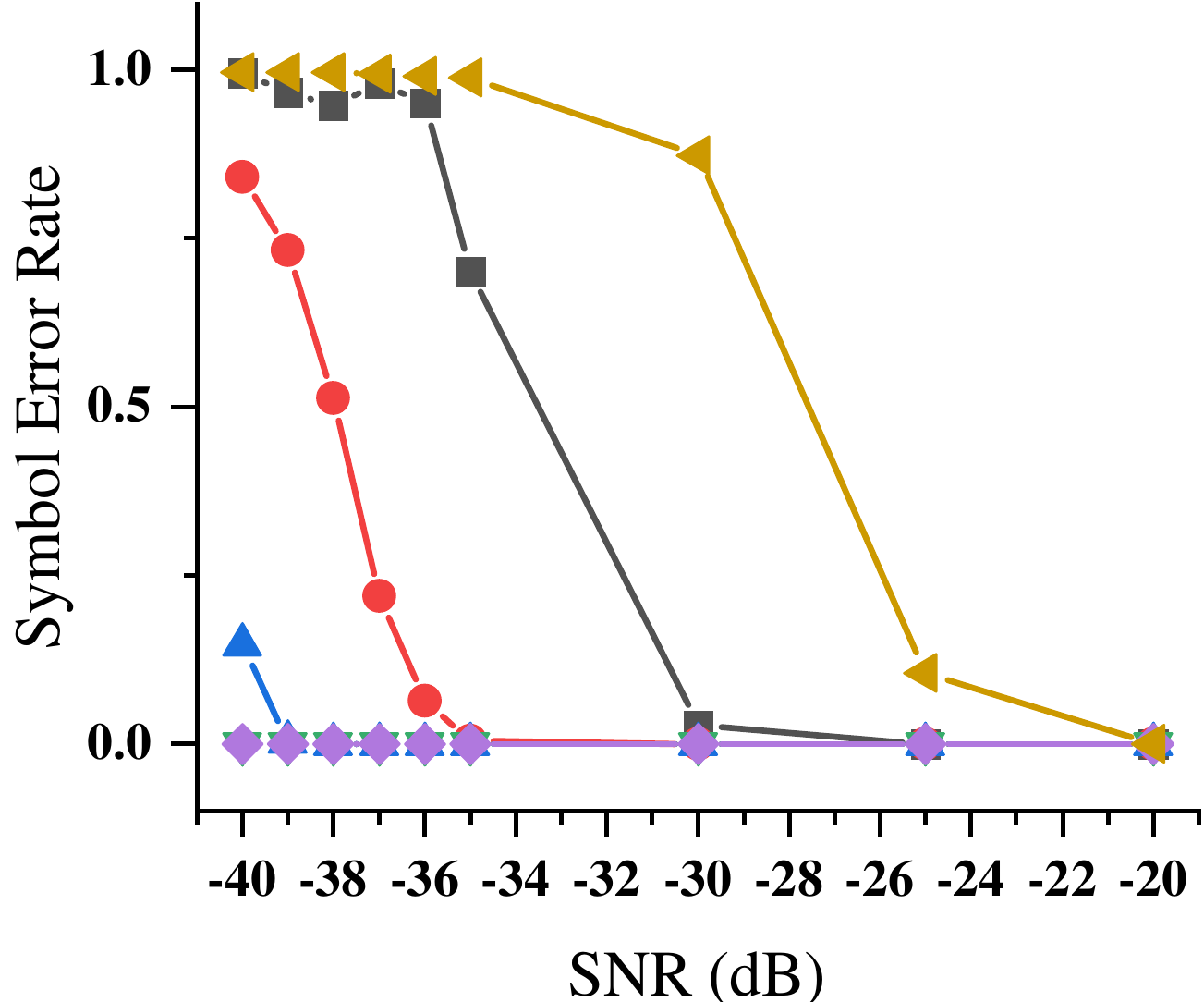}}
\\
\subfigure[B2-O  \label{SUMd}]{\includegraphics[width=0.327\columnwidth]{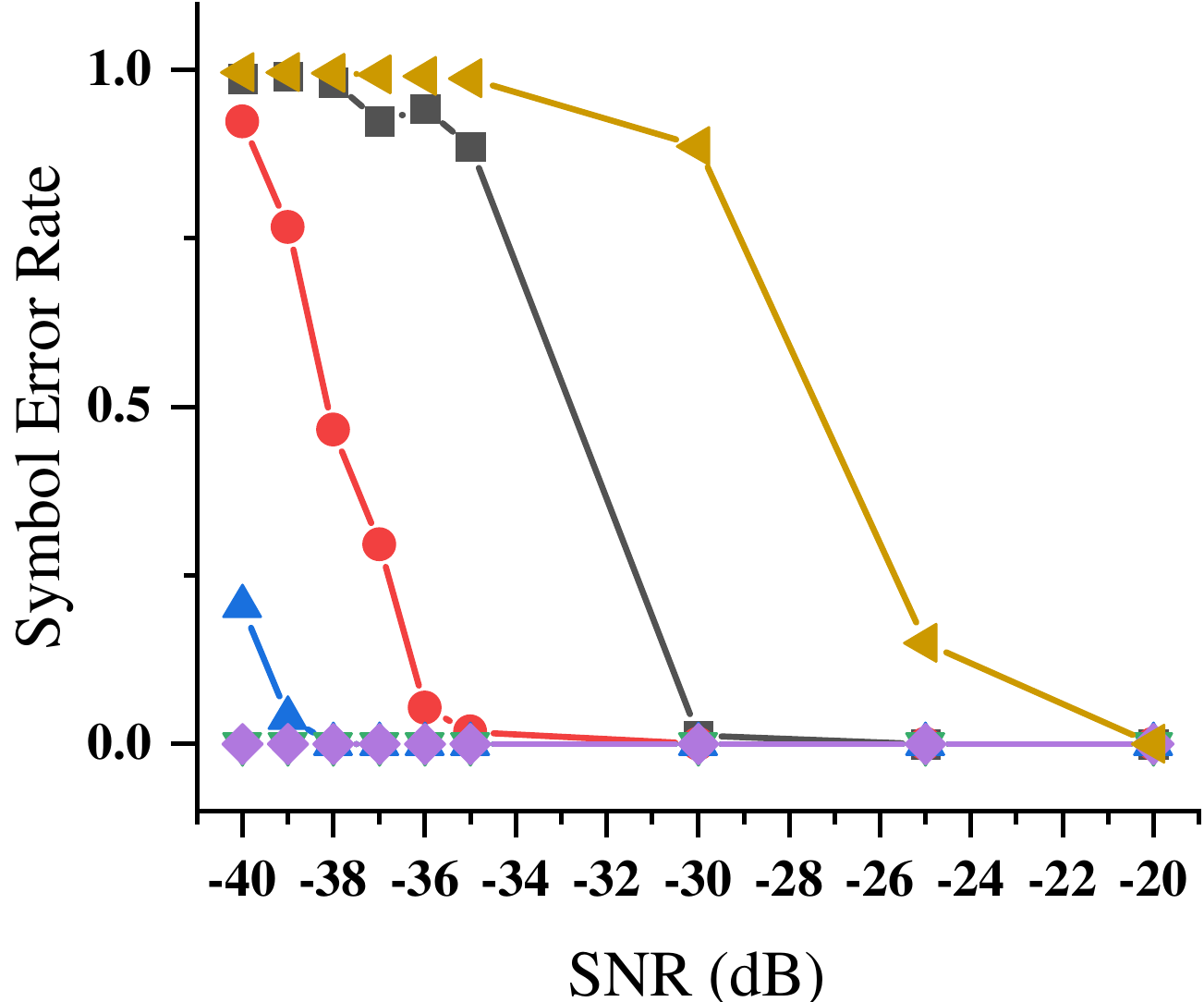}}
\subfigure[C1-O \label{SUMe}]{\includegraphics[width=0.327\columnwidth]{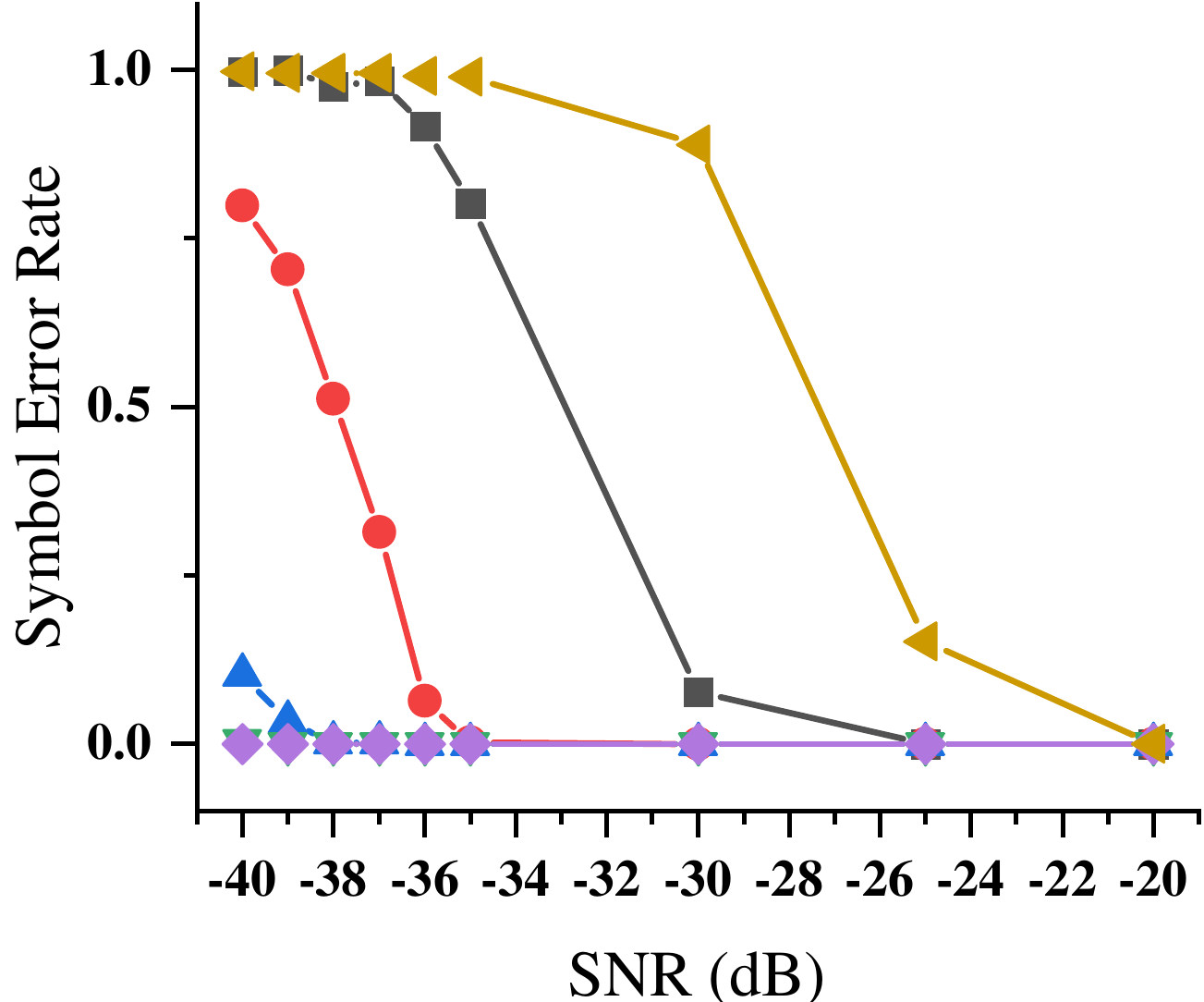}}
\subfigure[C2-O  \label{SUMf}]{\includegraphics[width=0.327\columnwidth]{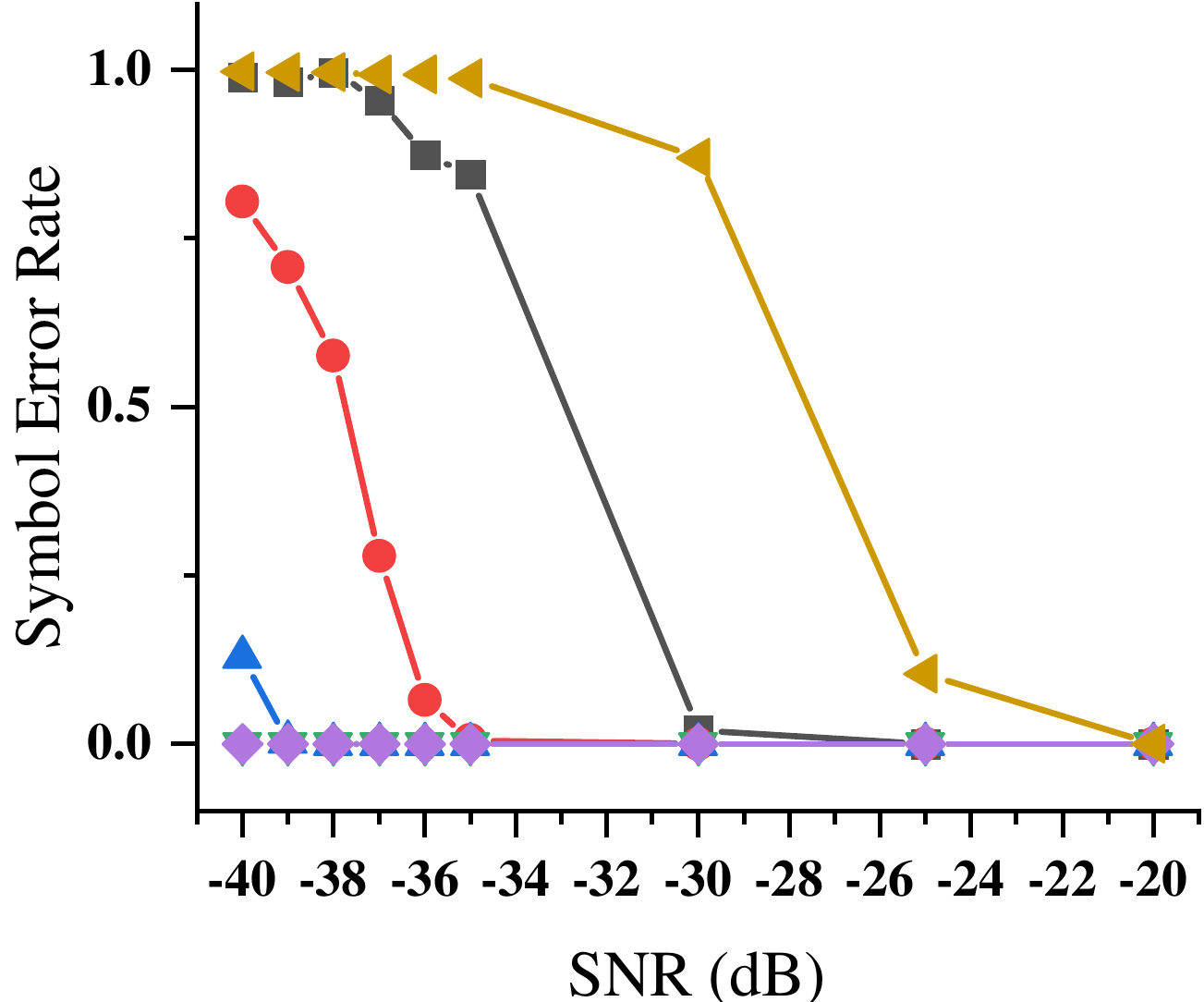}}
\\
\subfigure[D1-O  \label{SUMd}]{\includegraphics[width=0.327\columnwidth]{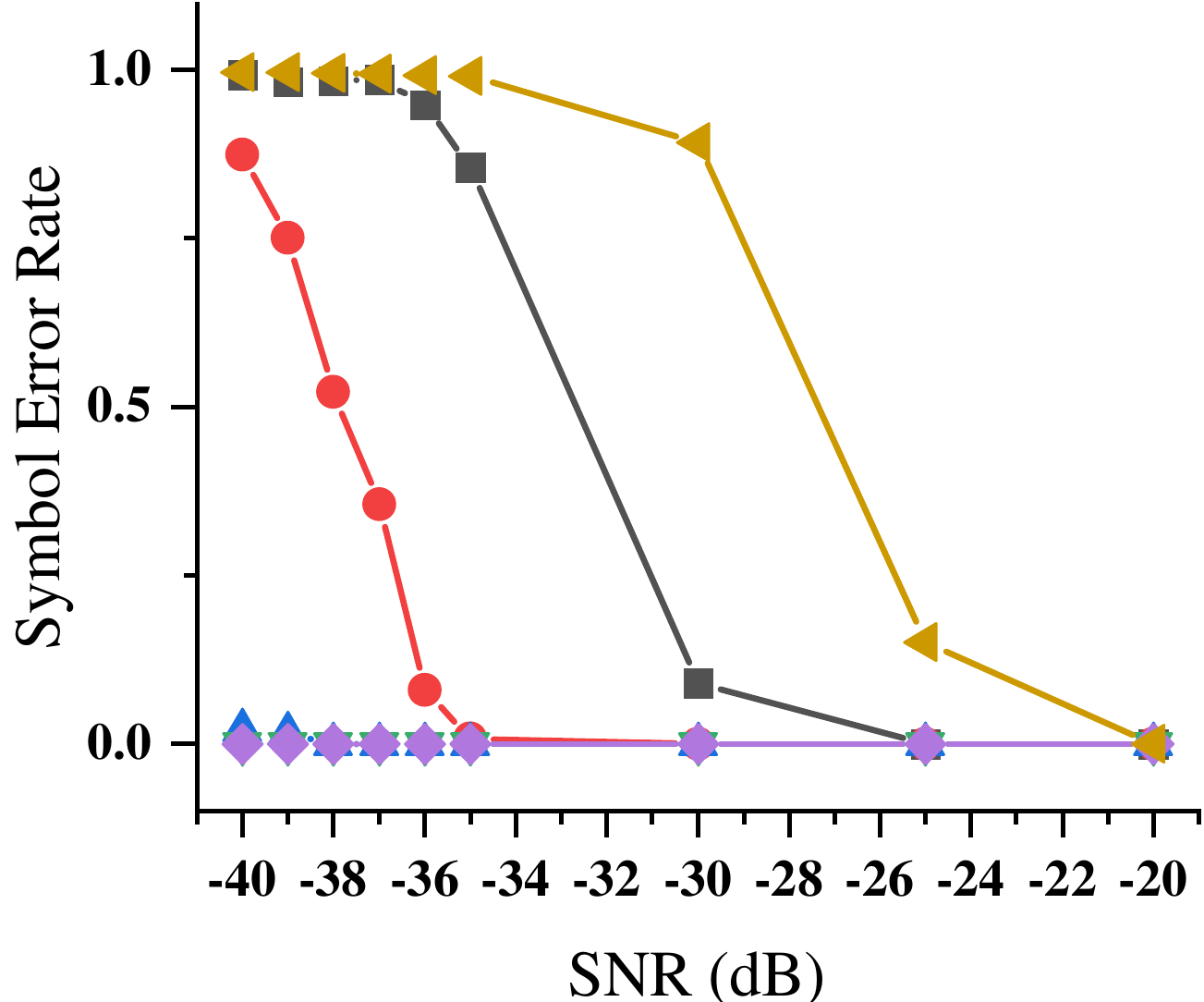}}
\subfigure[D2-O \label{SUMe}]{\includegraphics[width=0.327\columnwidth]{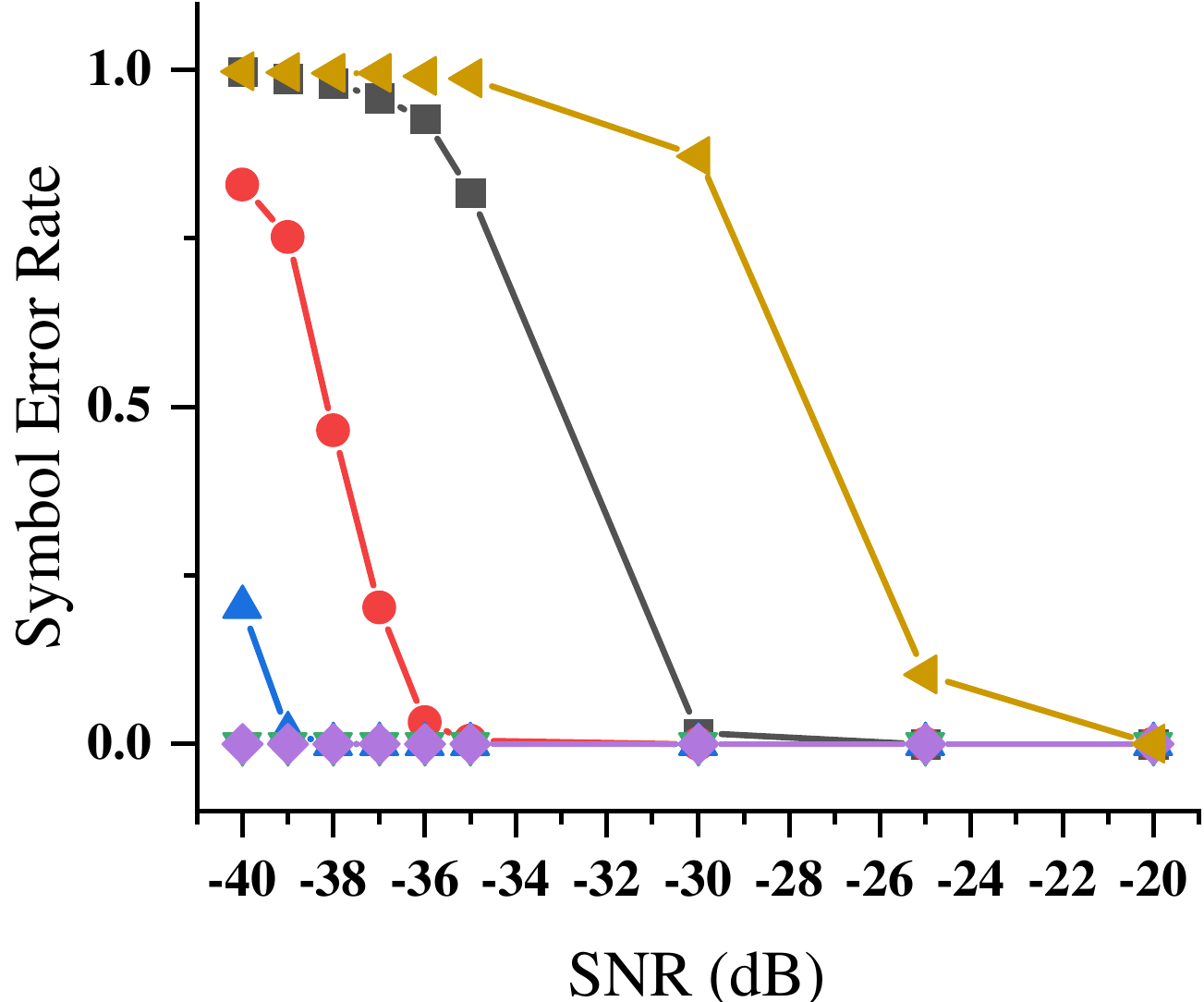}}
\subfigure[E1-O  \label{SUMf}]{\includegraphics[width=0.327\columnwidth]{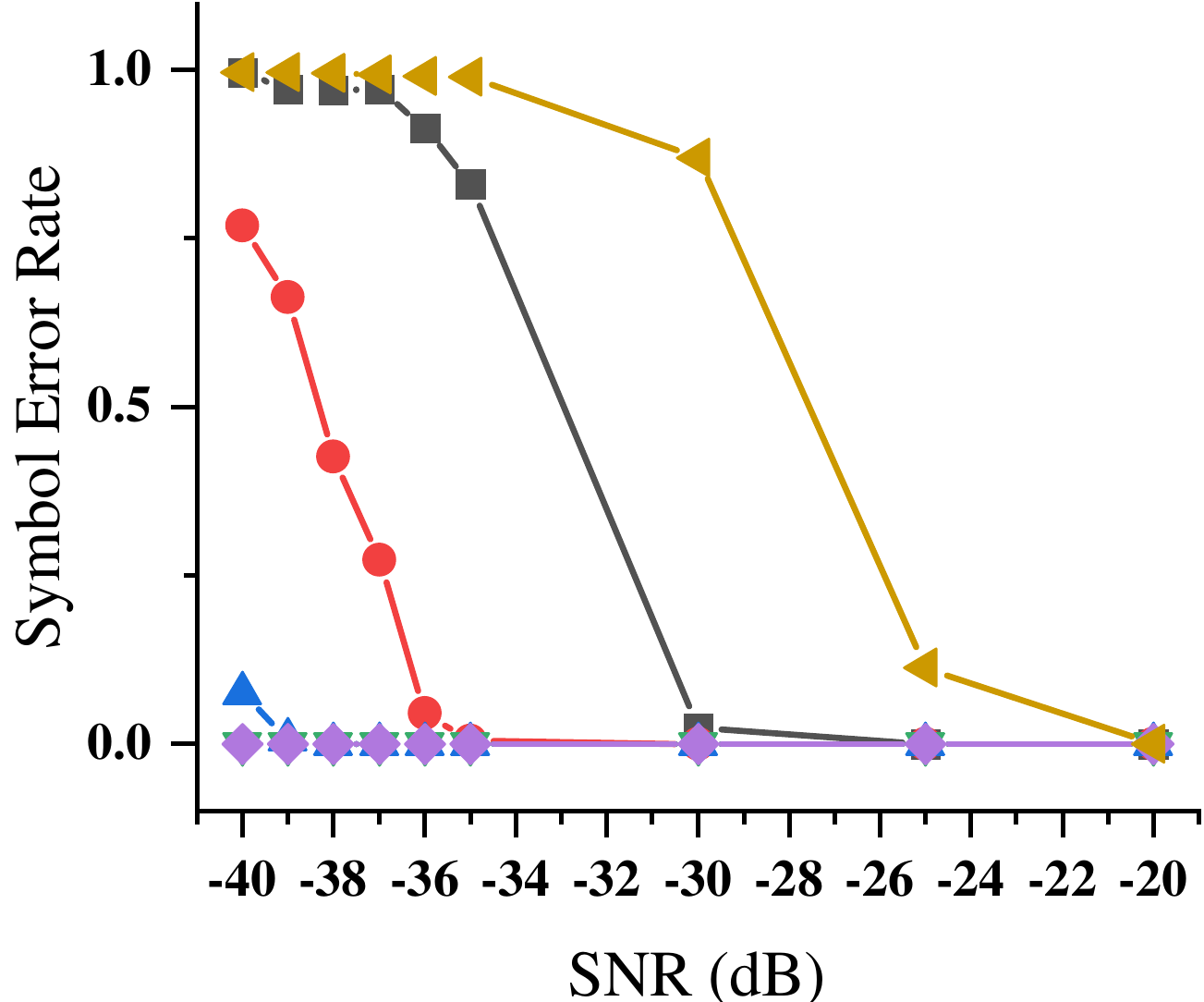}}
\caption{SNR Versus Symbol Error Rate for SF=13, with 7,000 runs per simulation point.}
\label{simres}
\end{figure}

Fig.~\ref{simres} shows the symbol error rate as a function of SNR for transmitters A1 to E1. At the chosen spreading factor (SF=13) and the parameters shown in Table~\ref{hhhh}, the results show that LoRa-Mod breaks down at an SNR of around -20 dB. However, LoRa-Mod-Enhanced continues to perform well up to -25 dB (Q=10), -35 dB (Q=100) and -39 dB (Q=500). The results are similar across all transmitter points  (E2 is not shown here but shares the same trend). Further simulations confirm that additional improvements can be achieved with higher spreading factors (at the cost of a decreased data rate). 

Longer Q's reduce the effective data rate, but the ability to adjust this provides flexibility. For example, more hostile channels can increase Q, effectively trading off data rate for improved robustness. We have simulated 10 transmitters operating simultaneously on a mixed LV-MV network, however, much like LoRa,  many more transmitters can operate simultaneously, sharing both time and frequency resources due to the orthogonality of the chirps at various combinations of SF and bandwidth. This is an important feature of the proposed scheme given the vast and sprawling nature of MV/LV networks, and the necessity for LV feeder load data (voltage and current) from all parts of the network.

\section{Hardware Implementation and Experimental Results}

Lora-Mod-Enhanced is implemented in Field Programmable Gate Array (FPGA) hardware using the high level architecture shown in Fig.~\ref{fpga}. The receiver is connected via UART to a Matlab App for visualisation and logging. The main features of the transmitter and receiver architectures are described next.

\subsection{Transmitter}
The FPGA based transmitter is based on a design first utilised in \cite{ROBSONupec2020}. The baseband complex chirp ($\Re$ and $\Im$) is stored in a pair of 32,708 point Read Only Memory (ROM) blocks. This is equivalent to an upscaling factor of 32 for a SF=10, 1,024 point symbol. Modulation is achieved by the ROM address counter, which allows the transmitter to output phase shifted versions of the complex chirp. A Numerically Controlled Oscillator (NCO) is used to generate a carrier frequency ($f_c$) for quadrature mixing of the complex chirp from the baseband to the passband. A 125 MSPS Digital to Analog Converter (DAC) is used to output the passband signal and amplification is provided by the OPA564 Power Operational Amplifier, which is capable of driving 1.5A at a gain-bandwidth product of 17 MHz. The LoRa bandwidth is 50 kHz.

\subsection{Receiver}
The receiver architecture shown in Fig.~\ref{RX} digitises the incoming signal using a 65 MSPS, 14-bit Analog to Digital Converter (ADC), operating at a lower sampling rate of 1.6 MHz. A Gaussian Noise generator provides the option of introducing an arbitrary level of Additive White Gaussian Noise (AWGN) to the incoming signal. The core has 16-bit resolution with a random distribution of $\pm 9.1 \sigma$ and a period of 2$^{176}$. Following downconversion and decimation by an FIR decimation filter (which downsamples the 1.6 MHz input signal by a factor of 32, down to the 50 kHz baseband), the dechirping process is carried out by a complex multiplier with a 1024-point ROM-based complex chirp.

An embedded processor is used to send the Lora-Mod output, $S$ and the Lora-mod-enhanced output, $\mathcal{H}$ via UART (baud = 115,200bps) to a Matlab app for visualisation and logging of the demodulated data.

To emulate a frequency selective channel, we have implemented a shift register and multiplier before the AWGN is added. This arrangements realises a simple 4-tap channel response with a separation of 4 LoRa samples between taps.

%
%
%


\begin{figure}
\centering
\includegraphics[width=\columnwidth]{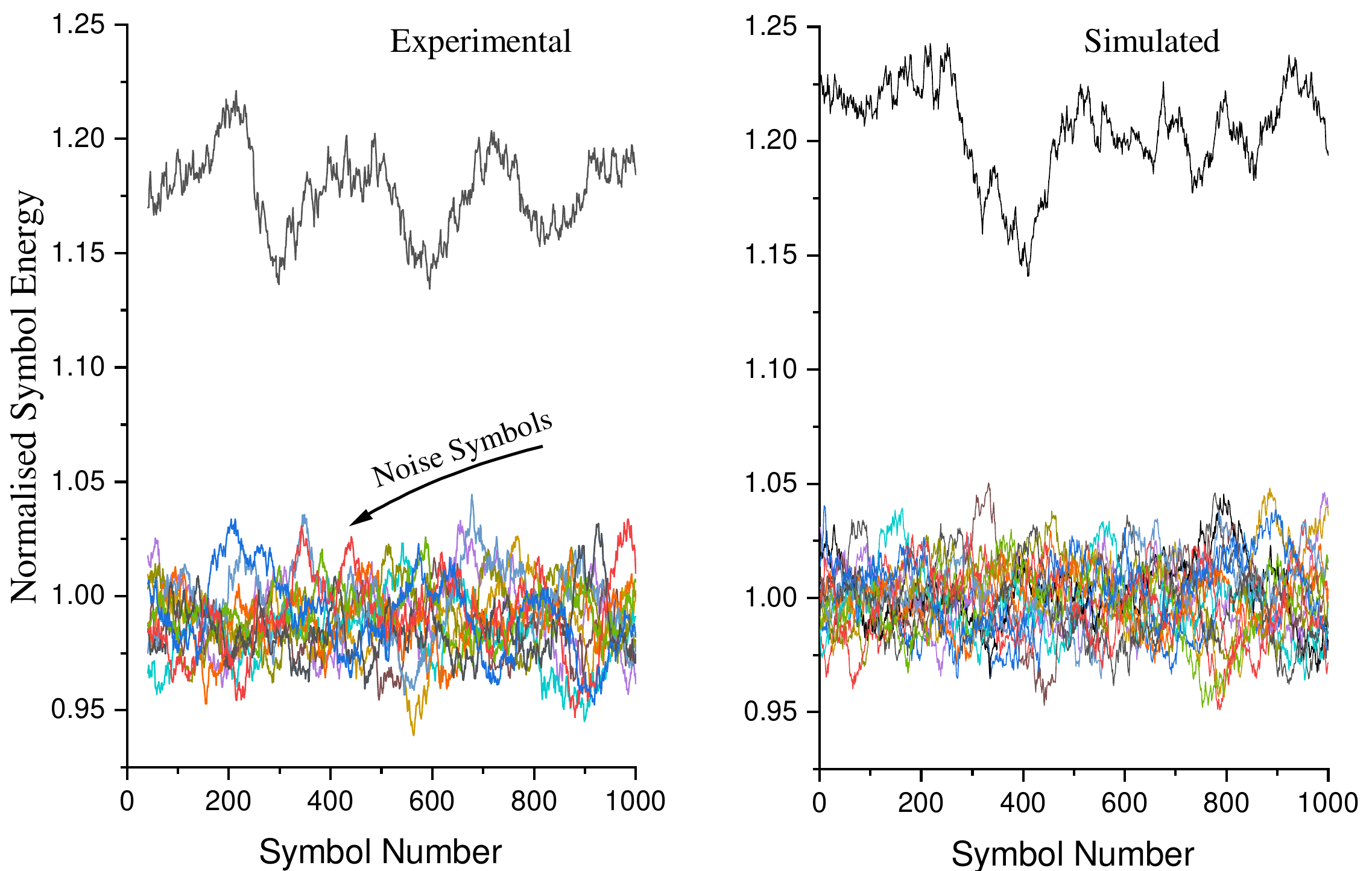} \\
\caption{Comparison of Experimental results, obtained with the FPGA prototype, and Simulation results using the same conditions. The AWGN is set to -19dB, SF=10.}\label{ep}
\end{figure}

\subsection{Experimental Results}
Fig.~\ref{ep} shows a comparison between experimental results and simulation using the same parameters. Here, an SNR of -19 dB is chosen, which, for SF=10, is below the threshold in which LoRa-Mod can operator. The plots show $\mathcal{E}$ and the noise symbols $S_\phi$ with a Q of 64. In this case, LoRa-Mod enhanced can achieve error-free communication with a time on air of 64$\cdot($2$^{SF})\cdot (\frac{1}{50,000})$ $\approx$ 1.3 s per running average. Good agreement between experimental and simulation is observed.

\section{Conclusion}

\begin{itemize}
\item A new PLC modulation scheme, based on a modification of the LoRa physical layer, has been proposed. This scheme has two versions i) LoRa-Mod, which subdivides the demodulated LoRa signal into a reduced set of `superbins', and ii) LoRa-Mod-Enhanced, which performs statistical averaging on each superbin.
\item The proposed scheme performs exceptionally well in the notoriously hostile LV-MV channel, coping with both low SNRs and extreme multipath. The condition of setting the superbin size to at least as great as the RMS delay spread is emphasised.
\item A new simulation methodology in the ATP-EMTP was developed, allowing millions of samples and numerous transmitters to be simulated simultaneously on a mixed (LV/MV) test network. The results demonstrate robust performance in extreme multipath and SNRs as low as -40 dB, with even better performance possible at higher spreading factors and longer moving averages.
\item The proposed scheme is implemented in FPGA hardware and experimental results match simulation.
\end{itemize}


%


\ifCLASSOPTIONcaptionsoff
  \newpage
\fi



\bibliographystyle{IEEEtran}
%

\bibliography{master2}
\end{document}